\documentclass[sensors,article,accept,pdftex,moreauthors]{Definitions/mdpi} 

\firstpage{1} 
\pubvolume{24}
\issuenum{21}
\articlenumber{6878}
\pubyear{2024}
\copyrightyear{2024}
\externaleditor{{Academic Editor: Alexios Mylonas} 
}
\datereceived{27 July 2024} 
\daterevised{9 October 2024} 
\dateaccepted{22 October 2024} 
\datepublished{26 October 2024} 
\hreflink{https://doi.org/10.3390/s24216878} 

\usepackage{pifont}
\newcommand{\cmark}{\textcolor{green}{\ding{51}}}%
\newcommand{\xmark}{\textcolor{red}{\ding{55}}}%
\usepackage{comment}
\usepackage{algorithm}
\usepackage{algorithmic}
\usepackage{subcaption}
\usepackage{svg}
\usepackage{array}
\newcolumntype{P}[1]{>{\centering\arraybackslash}p{#1}}
\newcolumntype{M}[1]{>{\centering\arraybackslash}m{#1}}
\newcolumntype{N}[1]{>{\arraybackslash}m{#1}}

\Title{CIPHER: Cybersecurity Intelligent Penetration-Testing Helper for Ethical Researcher}

\TitleCitation{CIPHER: Cybersecurity Intelligent Penetration-Testing Helper for Ethical Researcher}

\Author{{Derry Pratama} 
 $^{1}$\orcidA{}, Naufal Suryanto $^{2}$\orcidB{}, {Andro Aprila Adiputra} $^{1}$\orcidC{}, {Thi-Thu-Huong Le} $^{3}$\orcidD{}, \linebreak {{Ahmada Yusril Kadiptya}~$^{1}$\orcidE{}}, Muhammad Iqbal $^{1}$\orcidF{} and {Howon Kim} 
 $^{1,}$*\orcidG{}}

\AuthorNames{Derry Pratama, Naufal Suryanto, Andro Aprila Adiputra, Thi-Thu-Huong Le, Ahmada Yusril Kadiptya, Muhammad Iqbal and Howon Kim}

\AuthorCitation{{Pratama, D.;} 
 Suryanto, N.; Adiputra, A.A.; Le, T.-T.-H.; Kadiptya, A.Y.; Iqbal, M.; Kim, H.}

\address{%
$^{1}$ \quad School of Computer Science and Engineering, {Pusan National University, Busan 46241, Republic of Korea; } 
 derryprata@gmail.com (D.P.); androaprila@pusan.ac.kr (A.A.A.); yusril@pusan.ac.kr (A.Y.K.); iqbal@pusan.ac.kr (M.I.)\\
$^{2}$ \quad IoT Research Center, {Pusan National University, Busan 46241, Republic of Korea}; naufalsuryanto@gmail.com \\
$^{3}$ \quad Blockchain Platform Research Center, {Pusan National University, Busan 46241, Republic of Korea}; lehuong7885@gmail.com}

\corres{Correspondence: howonkim@pusan.ac.kr 

}

\abstract{Penetration testing, a critical component of cybersecurity, typically requires extensive time and effort to find vulnerabilities. Beginners in this field often benefit from collaborative approaches with the community or experts. To address this, we develop Cybersecurity Intelligent Penetration-testing Helper for Ethical Researchers (CIPHER), a large language model specifically trained to assist in penetration testing tasks \textcolor{black}{ as a chatbot}. 
Unlike software development, penetration testing involves domain-specific knowledge that is not widely documented or easily accessible, necessitating a specialized training approach for AI language models.
\textcolor{black}{CIPHER was trained} using over 300 high-quality write-ups of vulnerable machines, hacking techniques, and documentation of open-source penetration testing tools augmented in an expert response structure. Additionally, we introduced the Findings, Action, Reasoning, and Results (FARR) Flow augmentation, a novel method to augment penetration testing write-ups to establish a fully automated pentesting simulation benchmark tailored for large language models. 
This approach fills a significant gap in traditional cybersecurity Q\&A benchmarks and provides a realistic and rigorous standard for evaluating LLM's technical knowledge, reasoning capabilities, and practical utility in dynamic penetration testing scenarios.
In our assessments, CIPHER achieved the best overall performance in providing accurate suggestion responses compared to other open-source penetration testing models of similar size and even larger state-of-the-art models like Llama 3 70B and Qwen1.5 72B Chat, particularly on insane difficulty machine setups. This demonstrates that the current capabilities of general large language models (LLMs) are insufficient for effectively guiding users through the penetration testing process. We also discuss the potential for improvement through scaling and the development of better benchmarks using FARR Flow augmentation results.} 

\begin{document}




\section{Introduction}
\textcolor{black}{Penetration testing, a fundamental aspect of cybersecurity, involves probing computer systems~\cite{Denis2016PenetrationTC}, networks~\cite{Jayasuryapal2021ASO}, or web applications~\cite{Altulaihan2023ASO} to identify vulnerabilities that malicious actors could exploit. This proactive approach helps to uncover weaknesses before they can be manipulated, safeguarding sensitive data and maintaining system integrity. Despite its critical importance, penetration testing presents significant challenges, particularly for beginners. These challenges include understanding the complexity of modern IT infrastructures, mastering a wide range of tools, and gaining the necessary technical expertise to identify and exploit vulnerabilities effectively. Traditional penetration testing methods, which primarily rely on manual techniques, can be inefficient and labor-\mbox{intensive~\cite{mlonpentest}}, making them especially difficult for newcomers. Moreover, scaling these methods to meet the complexity and rapid evolution of modern IT systems further complicates the process. As cyber threats continue to grow in sophistication, there is a growing need for penetration testing methodologies that are not only efficient and comprehensive but also accessible to beginners, helping them overcome the steep learning curve.}

Automated tools and frameworks, such as Metasploit~\cite{metaesploit} and OpenVAS~\cite{openvas}, are designed to assist beginners by streamlining the penetration testing process. However, even with automation, users still need a certain level of expertise to effectively utilize these tools in dynamic pentesting situations, especially when interpreting and acting on the collected information.
Meanwhile, the application of large language models (LLMs) in penetration testing represents a cutting-edge area of research with promising results. Advanced models such as GPT-4~\cite{achiam2023gpt} have demonstrated significant potential to automate and improve various penetration testing tasks, including Linux privilege escalation and the identification of file-based vulnerabilities~\cite{happe2024llmshackersautonomouslinux}. However, implementing a straightforward LLM pipeline presents challenges, particularly in maintaining long-term memory for generating consistent and accurate commands throughout extended testing \mbox{sessions~\cite{deng2023pentestgpt}.}
Recent innovations like PentestGPT~\cite{deng2023pentestgpt} and AutoAttacker~\cite{xu2024autoattacker} have addressed these limitations. These systems leverage existing LLMs and open-source models, incorporating specialized frameworks to emulate human penetration testers' workflow and decision-making processes. While these advancements mark significant progress, it is important to note that these general-purpose models are not explicitly fine-tuned for the nuances of penetration testing. This lack of domain-specific optimization can potentially limit their effectiveness in handling complex, context-dependent scenarios often encountered in real-world penetration \mbox{testing environments}.

\textcolor{black}{Recognizing these limitations, we developed Cybersecurity Intelligent Penetration-testing Helper for Ethical Researchers (CIPHER) as a chatbot assistant specifically designed to support penetration testing tasks. CIPHER is a large language model fine-tuned on a specialized penetration testing dataset to assist ethical hackers by providing insights and recommendations. Trained on over 300 write-ups of vulnerable machines, hacking techniques, and open-source tools, CIPHER mimics expert reasoning and enhances the penetration testing process. It offers detailed explanations and guidance, particularly benefiting users with limited penetration testing experience. CIPHER is designed to be used specifically to guide beginner proceed with penetration testing a vulnerable machine for ethical research.}

\textcolor{black}{Building on this foundation, we introduced the Findings, Action, Reasoning, and Results (FARR) Flow, a novel method for the automated evaluation of large language models in penetration testing scenarios. FARR Flow acts as a structured benchmark, simulating realistic pentesting tasks to rigorously assess the model's technical knowledge, reasoning abilities, and practical effectiveness. By focusing on automated evaluation, FARR Flow provides a robust and realistic standard for measuring LLM performance in penetration testing, ensuring models are thoroughly tested in practical, real-world contexts.}

Our assessments indicate that CIPHER demonstrates a significant improvement in providing helpful responses for penetration testing over other models, highlighting the limitations of traditional cybersecurity evaluations in capturing the nuances of practical pentesting skills. Specifically, CIPHER showed a notable improvement, underscoring its potential to transform penetration testing practices. This paper presents several \mbox{key contributions}:
\textcolor{black}{\begin{itemize}
    \item The methodology for CIPHER's development, utilizing penetration testing write-ups structured as beginner- and expert-level conversations to train a large language model specifically for practical penetration testing assistance.
    \item A novel augmentation technique called Findings, Action, Reasoning, and Results (FARR) Flow that condenses these write-ups into structured formats for more efficient~evaluation.
    \item The development and evaluation of an automated pentesting simulation benchmark based on FARR Flow, providing a rigorous and realistic framework for evaluating LLM performance in practical penetration testing guidance.
    \item A comprehensive evaluation demonstrating CIPHER's effectiveness and superior performance over existing LLMs in providing accurate guidance in FARR Flow and MITRE ATT\&CK capabilities from PurpleLlama.
\end{itemize}
}

These contributions bridge the gap between AI advancements and practical cybersecurity applications, offering a new penetration testing domain-specific language model and reproducible benchmark for realistic penetration testing scenarios with consistent results.
 
\section{Background and Related Works}
\subsection{Large Language Models in Cybersecurity}
\subsubsection{Large Language Models} 

{Overview of Large Language Models:} 
Large language models (LLMs), exemplified by GPT-4, have revolutionized natural language processing through deep learning and extensive datasets, enabling them to understand and generate human-like text. These models predict the next word sequentially, capturing complex linguistic patterns and producing contextually relevant responses. Their applications include translation, summarization, and content creation, surpassing traditional NLP systems~\cite{vaswani2017attention}.

{Transformers:}
Introduced by Vaswani et al.~\cite{vaswani2017attention} in ``Attention is All You Need'' (2017), Transformers have reshaped NLP by employing self-attention mechanisms to process entire sequences concurrently. They excel at capturing long-range dependencies, outperforming sequential models like RNNs. Transformers comprise self-attention layers, feed-forward networks, and positional encoding to maintain word order. Variants such as BERT and GPT have set benchmarks in NLP and extended into computer vision tasks through Vision Transformers (ViTs).

{Open-Source LLMs:} 
Open-source LLMs like the Llama family by Meta AI (Toronto, Canada)~\cite{touvron2023llama}, Mistral (Paris, France)~\cite{jiang_mistral_2023}, Yi 1.5 by BAAI (Beijing, China)~\cite{ai_yi_2024}, and Qwen 1.5 by Alibaba Group (Hangzhou, China)~\cite{bai_qwen_2023} democratize access to powerful language models. Evolving from Llama 1 to Llama 3, these models increase token size and context length to enhance language understanding. Innovations like the mixture of experts in Mixtral by Mistral (Paris, France)~\cite{jiang_mixtral_2024} and unique training methods in Yi 1.5 push the boundaries further. Qwen 1.5 excels in multilingual contexts and incorporates agentic frameworks~\cite{qwen_team_qwen-agent_2024}, releasing smaller and powerful models~\cite{qwen_team_qwen15-moe_2024}, making them versatile tools~\cite{touvron2023llama,mann2020language}.

{Reasoning Enhancements in LLMs:}
LLMs benefit from enhancements like Chain of Thought (CoT) prompting and Self-Consistency (SC). CoT prompting encourages explicit reasoning steps, while SC verifies response correctness. Techniques like instruction fine-tuning (FLAN~\cite{wei_finetuned_2022}) and model mimicry (Orca~\cite{mukherjee_orca_2023}) further improve reasoning capabilities. Despite advancements, challenges in LLM reasoning persist, motivating ongoing \mbox{research~\cite{wei2022chain,achiam2023gpt}.}

{LLM-Based Chatbots:}
LLM-based chatbots like ChatGPT excel in customer support, education, and complex problem-solving by synthesizing large volumes of information into detailed responses. However, they lack the specialized knowledge required for offensive penetration testing~\cite{hassanin2024comprehensive}.

{Supervised Fine-Tuning:}
Supervised fine-tuning enhances model performance on domain-specific datasets, particularly in areas like penetration testing, ensuring accurate application of specialized language. Unlike Retrieval-Augmented Generation (RAG), fine-tuning improves the model's domain comprehension~\cite{gururangan2020don}.

{LLM Alignment:}
Efforts to align language models with human values focus on ensuring these models exhibit traits such as helpfulness and truthfulness. Reinforcement Learning with Human Feedback (RLHF) fine-tunes large language models (LLMs) to achieve this alignment. This process involves training a reward model to predict human preferences and then using algorithms like Proximal Policy Optimization (PPO) for fine-tuning. Although PPO generally yields better results, Direct Preference Optimization (DPO) simplifies the process by fine-tuning directly with human-labeled preferences~\cite{christiano2017deep, ouyang2022training}.

{Incorporating Domain-Specific Knowledge:}
Domain-specific knowledge enhances LLM accuracy in specialized fields like medicine~\cite{chen_meditron-70b_2023} and cybersecurity. Techniques such as Domain-Adaptive Pretraining (DAPT) and adaptive fine-tuning (AdaptLLM) are crucial for developing specialized models tailored for specific tasks, leveraging domain-specific datasets for improved insights~\cite{gururangan2020don, quan2024automatically}.

\subsubsection{Applications and Challenges in Cybersecurity}
In cybersecurity, large language models (LLMs) such as GPT-4~\cite{achiam2023gpt}, PentestGPT~\cite{deng2023pentestgpt}, hackingBuddyGPT~\cite{Happe_2023}, and HackerGPT~\cite{metta2024generative} have been deployed to assist with penetration testing and other security tasks. These models can analyze and synthesize large volumes of information, aiding in identifying vulnerabilities. However, using such advanced models poses significant challenges. The high cost and privacy risks associated with proprietary models like GPT-4, which handle sensitive vulnerability data, are significant concerns. Moreover, general-purpose LLMs often lack the specialized knowledge required for effective penetration testing.

\subsection{Existing Tools and Methodologies for Penetration Testing}
\subsubsection{Manual and Automated Tools}
\textcolor{black}{Manual tools such as Netcat, Wireshark, and Burpsuite Community Edition offer robust vulnerability discovery and exploitation frameworks. These tools, however, heavily rely on user expertise to interpret results and often require manual intervention to execute complex attack scenarios.}

\textcolor{black}{Meanwhile, automated tools have emerged to streamline the penetration testing process by automating vulnerability scanning and initial exploitation attempts. Examples including Metasploit~\cite{holik2014effective}, OpenVAS~\cite{aksu2019first}, and Nessus~\cite{kumar2014learning} use predefined algorithms to detect and exploit common vulnerabilities. While these tools enhance efficiency, they may overlook subtle vulnerabilities that require human intuition and context to identify.}

Combining manual and automated tools optimizes penetration testing effectiveness. Manual methods leverage human expertise to craft targeted attacks and interpret results, while automation accelerates routine tasks and broadens the scope of vulnerability detection. This hybrid approach ensures comprehensive security assessments addressing known vulnerabilities and emerging threats.

\subsubsection{Advancements in AI for Penetration Testing}
Recent advancements have seen the development of AI-driven tools designed to enhance penetration testing. \textcolor{black}{ Research efforts like PentestGPT~\cite{deng2023pentestgpt}, AutoAttacker~\cite{xu2024autoattacker}, hackingBuddyGPT~\cite{Happe_2023}, ReaperAI~\cite{valencia2024artificialintelligencenewhacker}, PenHeal~\cite{huang2024penhealtwostagellmframework}, Pentest Copilot~\cite{goyal2024hackinglazywayllm}, and  BreachSeek~\cite{alshehri2024breachseekmultiagentautomatedpenetration} leverage LLMs for various penetration testing tasks.} Despite their potential, these tools face several limitations:
\begin{itemize}
    \item {Context Loss and Memory Retention:} LLMs struggle with retaining long-term conversational memory, crucial for linking vulnerabilities across services to develop an exploitation strategy~\cite{deng2023pentestgpt,xu2024autoattacker}.
    \item {Testing Strategy Limitations:} LLMs often adopt a depth-first search approach, which may overlook other potential attack surfaces~\cite{deng2023pentestgpt}.
    \item {Inaccurate Operations and Commands:} LLMs may generate inaccurate commands or misinterpret outputs, leading to ineffective testing~\cite{fang2024llmweb,fang2024llmvuln}.
    \item {Limited Social Engineering and Image Interpretation:} LLMs lack capabilities in physical social engineering techniques and interpreting visual hints~\cite{xu2024autoattacker}.
    \item {Ethical Concerns:} The potential misuse of LLMs for automating attacks raises ethical issues~\cite{kumar2024ethics}.
\end{itemize}

\subsubsection{Leveraging LLMs for Penetration Testing}
There are distinct advantages to using native penetration testing LLMs versus general LLMs for cybersecurity tasks as shown in Table \ref{table:comparison_pentestingllm}. Native penetration testing LLMs offer superior performance, specialization, ease of use, efficiency, and security. These models are optimized explicitly for penetration testing tasks, resulting in higher accuracy and efficiency. Their deep specialization and out-of-the-box readiness make them more effective and more accessible to implement for specific penetration testing needs without extensive customization. Moreover, native models ensure better security by handling sensitive data locally, thus reducing the risk of data breaches.

On the other hand, general LLMs~\cite{achiam2023gpt} used as penetration testing agents~\cite{deng2023pentestgpt,Happe_2023,fang2024llmweb,fang2024llmvuln} provide greater versatility and scalability. These models can perform a broader range of tasks beyond penetration testing and can be scaled easily by integrating additional modules or upgrading to more advanced general models. However, they often require complex setups involving various frameworks, modules, and pipelines, which can be resource-intensive and challenging to manage. The increased complexity can introduce vulnerabilities and make troubleshooting more difficult. Without significant adjustments and configurations, general LLMs may not achieve the same performance and specialization as native penetration testing models.

\begin{table}[H]
\caption{{Comparison of current} 
 penetration testing LLMs research. Notes (\cmark: available; \xmark: not available).}
\newcolumntype{C}{>{\centering\arraybackslash}X}
\begin{tabularx}{\textwidth}{M{3.5cm}M{1.6cm}M{1.6cm}M{1.6cm}M{1.6cm}M{1.6cm}}
\toprule
\textbf{Name} & \textbf{Native LLM} & \textbf{Chat Assistant} & \textbf{Foothold} & \textbf{Priv. Escalation} & \textbf{Multiple Tools} \\
\midrule
LLM as Hackers~\cite{happe2024llmshackersautonomouslinux} & \xmark & \xmark & \xmark & \cmark & \xmark \\
\midrule
PentestGPT~\cite{deng2023pentestgpt} & \xmark & \cmark & \cmark & \cmark & \cmark \\
\midrule
AutoAttacker~\cite{xu2024autoattacker} & \xmark & \xmark & \cmark & \cmark & \xmark \\
\midrule
\textcolor{black}{hackingBuddyGPT}~\cite{Happe_2023} & \xmark & \xmark & \cmark & \cmark & \cmark \\
\midrule
\textcolor{black}{ReaperAI}~\cite{valencia2024artificialintelligencenewhacker} & \xmark & \xmark & \cmark & \cmark & \cmark \\
\midrule
\textcolor{black}{PenHeal}~\cite{huang2024penhealtwostagellmframework} & \xmark & \xmark & \cmark & \cmark & \cmark \\
\midrule
\textcolor{black}{Pentest Copilot}~\cite{goyal2024hackinglazywayllm} & \xmark & \cmark & \cmark & \cmark & \cmark \\
\midrule
\textcolor{black}{BreachSeek}~\cite{alshehri2024breachseekmultiagentautomatedpenetration} & \xmark & \cmark & \xmark & \cmark & \cmark \\
\midrule
LLM Agents Can Autonomously~\cite{fang2024llmweb,fang2024llmvuln} & \xmark & \xmark & \xmark & \cmark  & \xmark  \\
\midrule
\textcolor{black}{CIPHER (Ours)} & \cmark & \cmark & \cmark & \cmark & \cmark \\
\bottomrule
\end{tabularx}
\label{table:comparison_pentestingllm}
\end{table}

Given these considerations, we develop \textcolor{black}{CIPHER as a} native LLM tailored for penetration testing \textcolor{black}{chatbot assistant}. This approach ensures our model is optimized for the unique requirements of penetration testing tasks, thereby achieving higher accuracy and efficiency. A specialized model integrates domain-specific knowledge and tools more effectively, simplifying deployment and reducing the need for extensive customization. Additionally, handling sensitive data locally enhances security and minimizes the risk of exposure to vulnerable information.

\subsection{Specialized Cybersecurity Large Language Models}

\subsubsection{Development of Domain-Specific LLMs}
Specialized LLMs trained on domain-specific data, such as cybersecurity, are necessary to address the shortcomings of general-purpose models. Techniques like Domain-Adaptive Pretraining (DAPT) and adaptive fine-tuning improve performance by incorporating specialized knowledge and tools tailored for tasks like penetration testing~\cite{gururangan2020don, chen_meditron-70b_2023, lee2020biobert, cheng2023adapting, jiang2024improving}. These approaches enable models to understand and apply domain-specific language more effectively, enhancing their utility in specialized fields.

\subsubsection{Evaluation and Benchmarking Challenges}
Current evaluation frameworks for cybersecurity models, such as PurpleLlama CyberSecEval~\cite{bhatt_cyberseceval_2024} with MITRE ATT\&CK, primarily focus on predefined scenarios and known attack patterns~\cite{strom2018mitre}. These do not adequately measure a model's ability to discover novel vulnerabilities or adapt to dynamic situations encountered in real-world penetration testing. Previous works like PentestGPT and AutoAttacker lack reproducible benchmarks, relying on manual testing and human scoring, which introduces bias and limits \mbox{reproducibility~\cite{deng2023pentestgpt,xu2024autoattacker}}. Developing standardized, automated benchmarks that reflect real-world conditions remains a significant challenge for advancing the evaluation of cybersecurity LLMs.

\subsubsection{Reproducible Benchmark for Penetration Testing Scenario}
To address these challenges, we introduce the FARR Flow augmentation method, which automates the generation of penetration testing scenarios using high-quality write-ups. Unlike other existing penetration testing benchmarks for LLM, FARR Flow evaluation covers a wide range of tool usage capabilities, from reconnaissance to exploitation tools. FARR Flow evaluation results are also easily reproducible due to the scoring automation by using judge LLM. We also release our code on GitHub. This approach provides a dynamic and reproducible benchmark, assessing the LLM's ability to reason and act like an expert penetration tester.

As shown in Table \ref{table:comparison_benchmark}, our research demonstrates that FARR Flow evaluation fills the existing gap in current penetration testing benchmark for an LLM that is easily reproducible, open-source, and covers a wide range of tools.

\begin{table}[H]
\caption{{Comparison of penetration} 
 testing evaluation approaches. Notes (\cmark: yes; \xmark: not provided).}
\newcolumntype{C}{>{\centering\arraybackslash}X}
\begin{tabularx}{\textwidth}{M{4.5cm}M{2.5cm}M{3cm}M{2cm}}
\toprule
\textbf{Name} & \textbf{Multiple Tools} & \textbf{Consistent Reproducibility} & \textbf{Open Source Availability} \\
\midrule
LLM as Hackers~\cite{happe2024llmshackersautonomouslinux} &  \xmark & \cmark & \cmark \\
\midrule
PentestGPT~\cite{deng2023pentestgpt} & \cmark & \xmark & \xmark \\
\midrule
AutoAttacker~\cite{xu2024autoattacker} & \xmark & \cmark & \xmark \\
\midrule
\textcolor{black}{hackingBuddyGPT}~\cite{Happe_2023} & \cmark & \xmark & \cmark \\
\midrule
\textcolor{black}{ReaperAI}~\cite{valencia2024artificialintelligencenewhacker} & \cmark & \xmark & \cmark \\
\midrule
\textcolor{black}{PenHeal}~\cite{huang2024penhealtwostagellmframework} & \cmark & \cmark & \xmark \\
\midrule
\textcolor{black}{Pentest Copilot}~\cite{goyal2024hackinglazywayllm} & \cmark & \xmark & \xmark \\
\midrule
\textcolor{black}{BreachSeek}~\cite{alshehri2024breachseekmultiagentautomatedpenetration} & \cmark & \xmark & \cmark \\
\midrule
LLM Can Autonomously~\cite{fang2024llmweb,fang2024llmvuln} & \xmark & \xmark & \xmark \\
\midrule
\textcolor{black}{FARR Flow (Ours)} & \cmark & \cmark & \cmark \\
\bottomrule
\end{tabularx}
\label{table:comparison_benchmark}
\end{table}

\textcolor{black}{In Section \ref{farr-flow-reasoning-eval}, we propose FARR Flow augmentation and evaluation, which utilize existing high-quality penetration testing write-ups as a base for penetration testing flow. Therefore, creating the correct step of penetration testing of a specific vulnerable machine is no longer needed, enabling us to measure the performance of the model on existing vulnerable machines. Unlike traditional human scoring, which cannot be consistently reproducible, in FARR Flow we can  change the evaluator model to avoid and minimize bias using multiple SOTA models as evaluators.}

\section{Methodology}
\textcolor{black}{CIPHER development focused on two key contributions, the first is where we develop our specific chatbot assistant model that is highly focused on suggesting the correct next step to be taken based on currently obtained information. Secondly, we provide an automated benchmarking standard for measuring the penetration testing advice from the LLM that is easily reproducible and has consistent results.}

\textcolor{black}{Unlike other penetration testing LLMs that rely solely on Retrieval-Augmented Generation (RAG) or agent-based pipelines dependent on the OpenAI model, CIPHER's knowledge is primarily drawn from highly curated penetration testing write-ups. These resources provide a rich foundation of both successful and failed penetration tests, offering invaluable insights into the reasoning and decision-making processes of experienced penetration testers. We believe that merely relying on the literature is insufficient for teaching a model the nuances of penetration testing; the real expertise comes from understanding how professionals adapt and learn through hands-on experience.}

\begin{figure}[H]
\includegraphics[width=.98\linewidth]{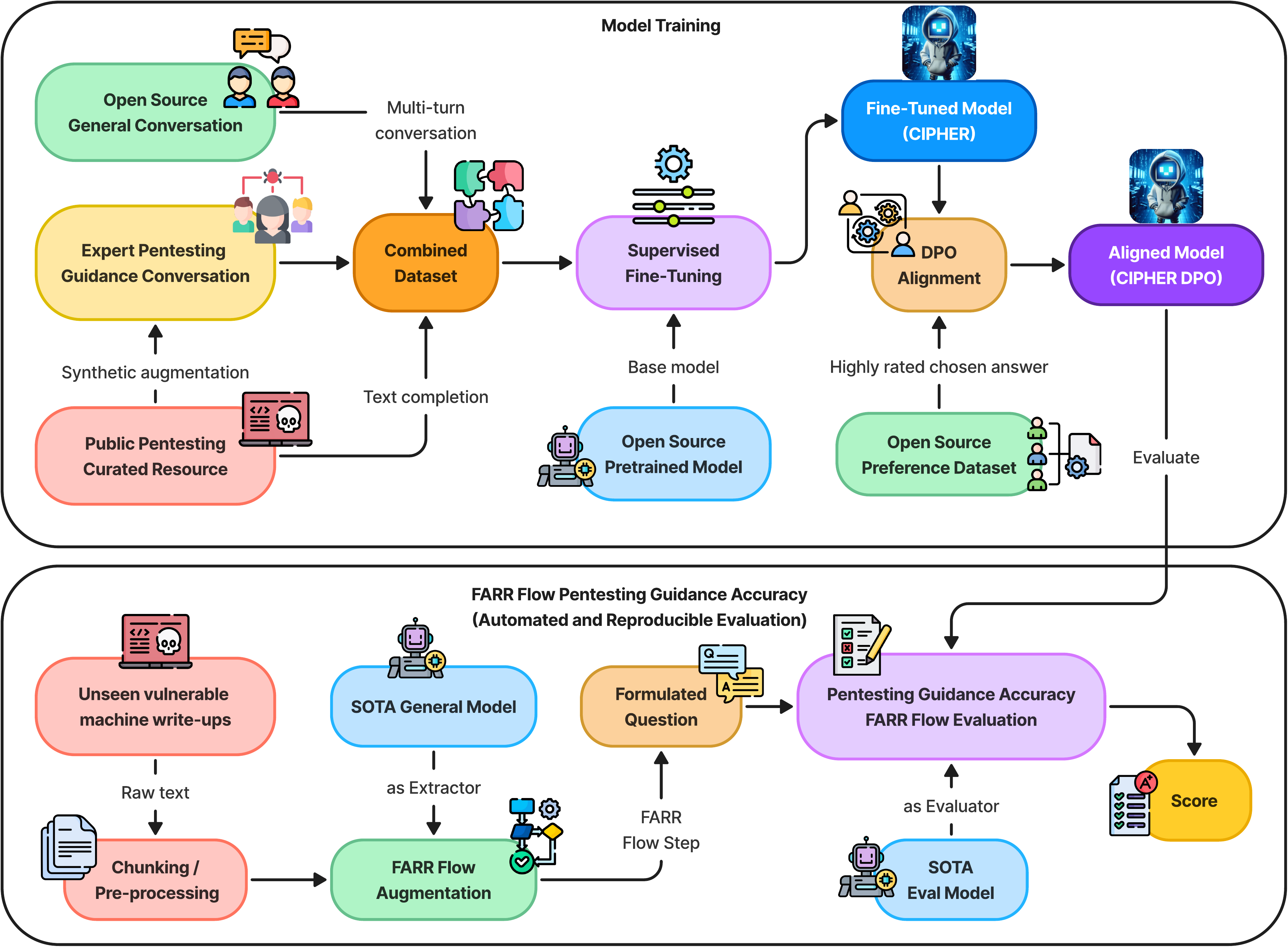}
\caption{\textcolor{black}{{CIPHER development} 
 methodology overview for training and novel automated evaluation.}}
\label{fig:cipher-methodology} 
\end{figure}

\textcolor{black}{Building on this core, CIPHER's methodology {Figure} 
 \ref{fig:cipher-methodology} leverages publicly available penetration testing resources and write-ups to form the backbone of its knowledge base. The curated mixture of resources, detailed in Table \ref{table:pentest_mixture_dataset}, ensures that CIPHER learns from real-world scenarios that capture both successes and failures. This knowledge is further augmented with expert-guided conversations (explained in Section \ref{section:Augmentation}), which help the model understand the context and decision-making paths within the penetration testing process. Additionally, to retain a well-rounded conversational ability, CIPHER integrates general world knowledge through resources like OpenHermes 2.5~\cite{OpenHermes2.5}, ensuring it can handle both technical penetration testing tasks and general discussions as a chatbot.}

\textcolor{black}{To further refine CIPHER's capabilities, a supervised fine-tuning process is performed using the Axolotl framework~\cite{githubGitHubAxolotlaicloudaxolotl}, enabling quick and reproducible training with hyperparameters detailed in Section \ref{section:supervised-fine-tuning}. Following fine-tuning, an alignment process is conducted using openly available preference datasets to ensure that CIPHER aligns with human preference, as discussed in Section \ref{section:dpo}.}

\textcolor{black}{
For evaluation, we utilize the latest, unseen vulnerable machine write-ups as references. These write-ups are processed into chunks and extracted using FARR Flow augmentation, which integrates the strengths of a state-of-the-art (SOTA) general LLM model, described in Section \ref{farr-augmentation}. Once these flows are gathered, we evaluate CIPHER’s performance by formulating questions based on the extracted knowledge, as outlined in Section \ref{farr-flow-reasoning-eval}.
}
\subsection{Architecture Design}

\begin{figure}[H]
\includegraphics[width=.98\linewidth]{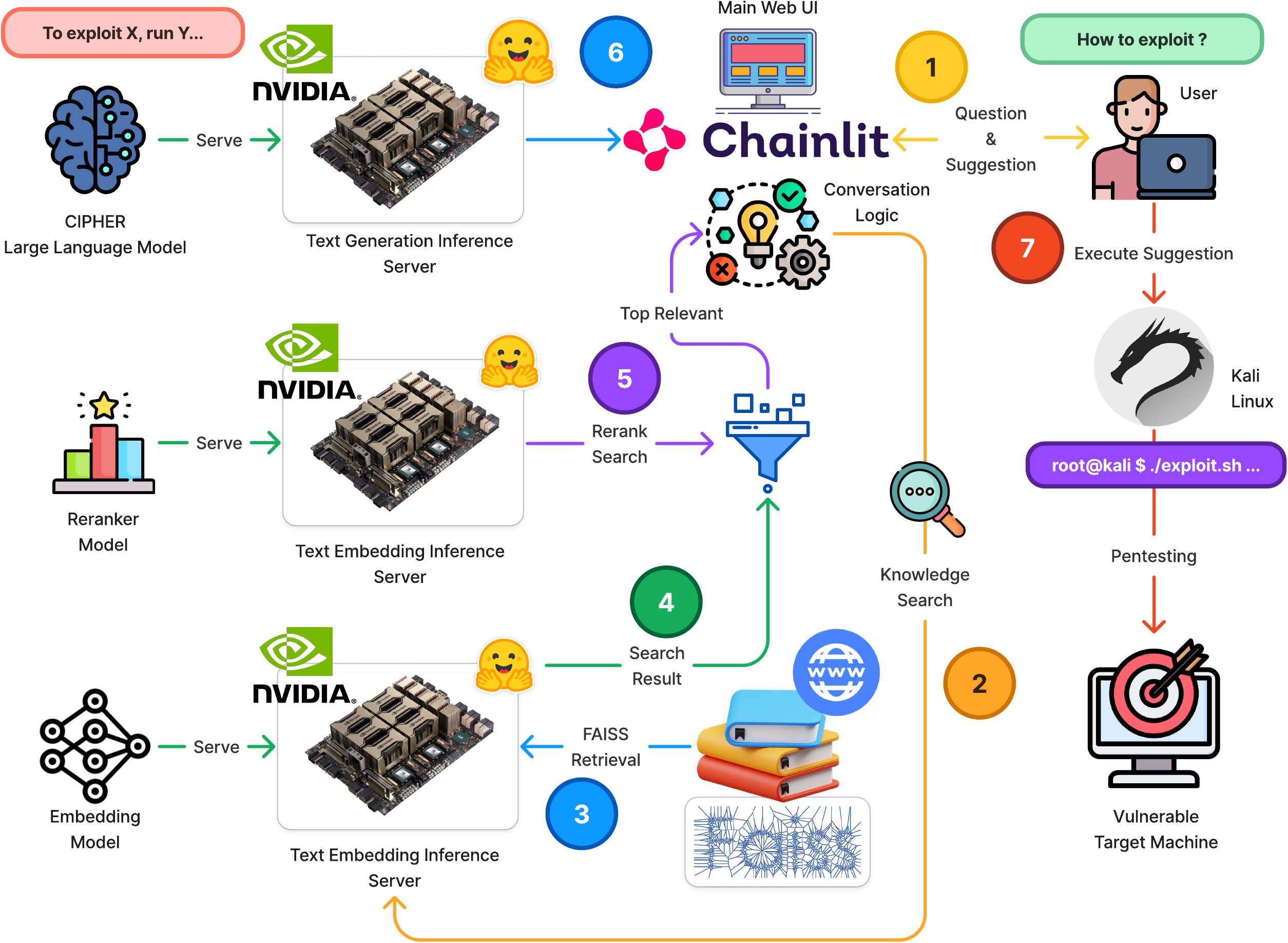}
\caption{{The architecture} 
  of CIPHER deployed as chat assistant. (1) The user submits a query and known information, which is converted into text embeddings (2). FAISS VectorDB performs cosine similarity matching with the knowledge database (3). The reranker filters and reorders results based on relevance (4). The top-ranked response is then used to build reference (5), leading to the generation of the best suggestion for penetration steps (6). The user executes this suggestion on the attacker machine (7).} 
\label{fig:cipher-arch}
\end{figure}

The main architecture of CIPHER is shown in Figure \ref{fig:cipher-arch}. CIPHER's primary objective is to simplify identifying system vulnerabilities by providing expert guidance throughout the penetration testing workflow.
Recognizing the challenges beginners face in using specialized tools and grasping core concepts, CIPHER offers the following:
\begin{itemize}
    \item Explanation of the question.
    \item Intuitive, step-by-step instructions with reasoning.
    \item Insightful response that resembles expert penetration tester reasoning.
    \item Additional practical examples for the action.
\end{itemize}

\textcolor{black}{In order to enhance the accuracy in suggesting command line usage on deployment, we added advanced Retrieval Augmented Generation (RAG)~\cite{lewis_retrieval-augmented_2020} in full architecture as seen in Figure \ref{fig:cipher-arch}. Initially, user question will be processed by the mxbai-embed-large-v1 embedding model with 335M parameters~\cite{sean_lee_open_2024, li_angle-optimized_2023} to find the similar hacking technique and command line documentation. Then similar documentation will be reranked using reranker model~\cite{xiao_bge-reranker-large_2024, xiao_c-pack_2024} to find the best document chunks. CIPHER will use the related chunks as in-context learning~\cite{dong2024surveyincontextlearning} material to answer the user question with a suggestion as accurate as possible. CIPHER utilizes BGE-reranker-base (278M parameter) that came from BGE family embedding model~\cite{xiao_c-pack_2024}.  The embedding model will convert the text into vectors to convert all pentesting-related content. To query the content, we retrieve top-3 closest content based on the inputted prompt using Cosine Similarity to reduce unrelated content. This research will discuss the core development of the language model to support this environment architecture.}

CIPHER focuses on facilitating penetration testing tools and emphasizes developing the user's ability to reason and make informed decisions during the testing process. By offering detailed explanations and reasoning for each step, CIPHER helps users understand the underlying principles and methodologies experienced penetration testers use. This dual focus on tool usage and expert reasoning equips users with the skills and confidence to conduct thorough and effective penetration tests, ultimately enhancing the security of their systems.

The development of CIPHER aims to bridge the gap between novice and expert, equipping users with the technical skills and reasoning abilities necessary to improve their system security posture significantly.

\subsection{Dataset}
Developing a pentesting chat assistant requires two fundamental capabilities: general chat assistance and domain-specific pentesting expertise. We utilize distinct datasets to address each of these capabilities.

\subsubsection{General Assistant Capability} \label{section:general-assistant-capability}
We leverage the OpenHermes 2.5 dataset for our general assistant capability, which is currently recognized as one of the best open-source conversation collections. This dataset was originally used to fine-tune the OpenHermes 2.5 model sourced from various origins, including Orca, Glaive, and other GPT-4 responses in diverse questions and \mbox{instructions~\cite{OpenHermes2.5}}.  When used to fine-tune a base model, this conversational dataset has improved performance on multiple general-use benchmarks such as MMLU~\cite{hendrycks_measuring_2021}, HumanEval~\cite{chen_evaluating_2021}, and \mbox{TruthfulQA~\cite{lin_truthfulqa_2022}.}

Note that this dataset primarily enhances the model's ability to formulate responses rather than expanding its knowledge base. The dataset provides examples of responding to greetings, questions, and instructions, effectively teaching the model appropriate ``answering styles.'' The varying performance of models fine-tuned on this dataset suggests that the underlying knowledge comes from pretraining rather than from examples of \mbox{answering questions}.

The dataset already includes patterns for responding to user greetings, questions, and instructions. However, the effectiveness of this dataset in improving model performance can vary depending on the base model used, further supporting the notion that true knowledge stems from pretraining rather than from examples of question-answering.

The OpenHermes 2.5 dataset is a comprehensive collection comprising over 1.6 million entries, designed to cover a wide range of conversational scenarios and specialized knowledge domains. Key components include the following (with rough estimations):
\begin{itemize}
\item Creative prompts from Airoboros 2.2 (44.8 K);
\item Domain-specific expertise from CamelAI in MATH, Physics, Chemistry, and Biology (50 K, 20 K, 20 K, and 20 K, respectively);
\item Real-world complex prompts from Chatbot Arena (7 K);
\item Up-to-date data from Collective Cognition (as of {9 November 2023}) 
 (156);
\item Chain-of-thought examples from the Alpaca dataset, produced by GPT-4 (52 K);
\item Evolving complexity prompts from Evol Instruct (70 K and 140 K versions);
\item Tool usage scenarios from Glaive Code Assistant (136 K);
\item \textls[-15]{General-purpose conversational data from GPT4-LLM (54.6 K) and GPTeacher \mbox{(35.5 K)}};
\item Specialized task-oriented datasets like Medical Tasks (4.6 K) and MetaMath (40 K);
\item Reasoning-focused entries from SlimOrca (550 K) and Platypus (24.9 K);
\item Real conversation enhancements from ShareGPT (356 K);
\item Versatility-boosting prompts from Unnatural Instructions (9 K).
\end{itemize}

This rich mixture of datasets aims to create a well-rounded foundation for the AI assistant, covering various aspects of knowledge, reasoning, and interaction styles. The diversity of sources enhances the model's ability to handle various queries and \mbox{tasks effectively}.

\subsubsection{Penetration Testing Capability}
Domain-specific knowledge enhancement has been extensively \mbox{researched~\cite{cheng2023adapting,jiang2024improving,chen_meditron-70b_2023}.} The most effective approach to improve domain knowledge is to enrich the corpus with domain-specific text. Our focus is on basic cybersecurity and penetration \mbox{testing knowledge.}

Penetration testing is a broad field that requires a strong foundation before identifying specific services or systems vulnerabilities. A penetration tester must understand how various services and systems operate and common vulnerabilities. While most public internet data provide basic knowledge of systems, which is reflected in current LLMs, the same is not true for vulnerability data and exploitation techniques. Due to safety concerns, most LLMs are biased towards protecting vulnerabilities or preventing exploitation techniques rather than suggesting how to exploit them.

To focus CIPHER's knowledge as a red-teamer, we scraped high-quality and frequently referenced hacking techniques, as shown in Table \ref{table:pentest_mixture_dataset}. This includes open public collections of OSCP notes and popular GitHub repositories. We incorporated compact documentation of popular tool command-line usage, using Cheatsheets~\cite{Lane2024cheat} and TLDR~\cite{githubGitHubTldrpagestldr} to improve general command-line argument knowledge. The Kali Tools~\cite{kaliKaliTools} dataset strengthens the model's knowledge of penetration testing tools in Kali Linux, which we chose as the target environment for CIPHER. The Identity dataset augments CIPHER's self-awareness based on its purpose and specifications.

\begin{table}[H]
\caption{Penetration testing essentials mixture.}
\newcolumntype{C}{>{\centering\arraybackslash}X}
\begin{tabularx}{\textwidth}{p{4cm}p{7cm}}
\toprule
\textbf{Category} & \textbf{Content} \\
\midrule
Fundamental Knowledge & Cheatsheets~\cite{Lane2024cheat}, TLDR~\cite{githubGitHubTldrpagestldr}, Identity, Kali tools~\cite{kaliKaliTools} \\ \midrule
Pentesting Knowledge & OSCP Notes~\cite{gitbookgabbarIntroductionOSCP}, OSCP Playbook~\cite{fauzi_oscp_playbook} \\ \midrule
Privilege Escalation & GTFOBins~\cite{gtfobinsGTFOBins}, LOLBAS~\cite{lolbasprojectLOLBAS} \\ \midrule
Hacking Techniques & Hacktricks~\cite{hacktricksHackTricksHackTricks}, PayloadAllTheThings~\cite{githubGitHubSwisskyrepoPayloadsAllTheThings} \\ \midrule
Practical Knowledge & 0xdf Hack The Box Write-ups~\cite{0xdf0xdfHacks} \\ \bottomrule
\end{tabularx}
\label{table:pentest_mixture_dataset}
\end{table}

We added advanced penetration testing knowledge from Gabbar OSCP notes~\cite{gitbookgabbarIntroductionOSCP} and OSCP Playbook~\cite{fauzi_oscp_playbook}, chosen for their structured and high-quality content. For privilege escalation, we included GTFOBins~\cite{gtfobinsGTFOBins} (Linux) and LOLBAS~\cite{lolbasprojectLOLBAS} (Windows), considering misconfiguration as a common attack vector.

To deepen CIPHER's understanding of hacking and exploitation techniques, we incorporated the open-source Hacktricks~\cite{hacktricksHackTricksHackTricks} books, covering reconnaissance to undocumented exploitation techniques. We further enhanced exploitation skills by adding PayloadAllTheThings~\cite{githubGitHubSwisskyrepoPayloadsAllTheThings}, which provides a rich collection of payload examples and \mbox{usage instructions.}

While fundamental knowledge, hacking techniques, and payloads can equip a medium-level penetration tester, achieving objectives with limited resources is the core challenge. This requires understanding situational context, prioritizing attack vectors, and developing intuition through experience. Such knowledge is rarely documented and is typically gained through practice.

To address this, we utilized a collection of high-quality write-ups from over 300 Hack The Box~\cite{hacktheboxHackBox} machines, sourced from the 0xdf blog~\cite{0xdf0xdfHacks}. We selected these write-ups for their expert-level content and text-based format, which facilitates straightforward processing as a dataset. The write-ups provide detailed descriptions of visual elements, ensuring that the content remains comprehensive even without accompanying images.

\textcolor{black}{While there are other collections, such as those from TryHackMe~\cite{tryhackmeTryHackMeCyber} or \mbox{VulnHub~\cite{vulnhubVulnerableDesign}}, we found that the quality and quantity of the 0xdf blog's write-ups surpass these alternatives. One potential downside of this decision is the introduction of bias, either towards Hack The Box machines or the author's methodology. However, this is not a significant issue, as Hack The Box is widely regarded as a leader in cybersecurity skills and training platforms. These are the currently best available data to use.}

\subsubsection{Augmentation} \label{section:Augmentation}
We process the dataset through several steps:

{(1) Raw Context Dataset}:  
For open-source datasets available in markdown format, no post-processing is needed. Web-formatted datasets are converted to markdown to preserve content structure. We split the datasets into chunks of 500, 1 K, and 2 K tokens, maintaining parent headers for nested sections to preserve context. This approach, providing the same information at different lengths, aids in better model generalization.

{(2) Conversation Dataset}: 
While raw context datasets expand domain-specific knowledge like DAPT~\cite{gururangan2020don}, they do not fully prepare the model to answer and suggest like an expert. We augment and prioritize Hack The Box write-ups of vulnerable machines into conversations, simulating exchanges between novices and experts. While conversations do not add new knowledge, they help the model mimic expert answering styles, reducing mistakes and improving response quality~\cite{mukherjee_orca_2023}.

\begin{figure}[H]
       \includegraphics[width=1\linewidth]{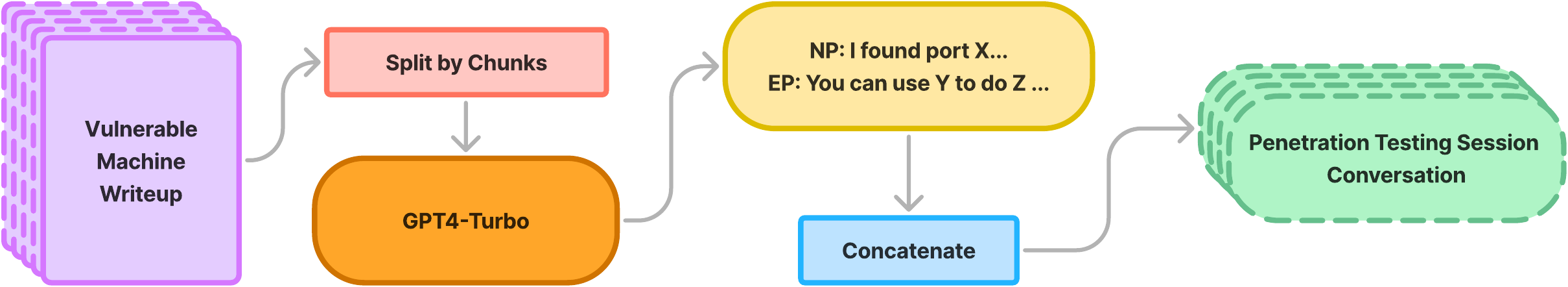}
    \caption{Each machine write-up is chunked into smaller pieces to extract multiple conversations, concatenated into a complete penetration testing session dialogue.}
    \label{fig:gen-conv-pipeline}
\end{figure}

{(3) Conversation Generation Pipeline}:
We designed CIPHER to assist with penetration testing tasks, focusing on scenarios where a novice reports findings and situations and an expert provides suggestions, definitions, reasoning, and examples for the next steps. As seen in Figure \ref{fig:gen-conv-pipeline}, our pipeline generates conversations from 500-token chunks of raw text, ensuring focused discussions on specific topics instead of broader context.

{(4) Self-sufficient and Generalized Conversation}: 
Table \ref{table:prompt_conv_gen} shows our prompt for generating synthetic conversations. We emphasize \textbf{{self-sufficient}} and \textbf{{generalized}} conversations to create independent scenarios that convert write-ups into current situations, avoiding direct references to the source material.

\begin{table}[H]
\caption{{Prompt for conversation} 
 generation, with concise instructions where each word is crucial for generation quality.}
\setlength{\tabcolsep}{4mm}  
\begin{tabularx}{\textwidth}{p{13cm}}
\toprule 
\textbf{Prompt Format} \\
\midrule 
<500 tokens of write-up chunk>

---

Convert the write-up above into a self-sufficient generalized conversation without referring to this context.

The conversation is a question from a novice pentester and a helpful answer from an expert pentester.

The newbie always asking what to do next.

The experts always provide reasoning explanations, then followed by examples.

The conversation is multiple turns, step by step. \\
\bottomrule
\end{tabularx}
\label{table:prompt_conv_gen}
\end{table}

{(5) Newbie and Expert Roles}: 
We explicitly define Newbie and Expert roles to generate beginner-level questions and helpful, detailed expert responses.

{(6) Next Step Prediction}: 
We incorporate questions like ``What should I do?'' or ``What is the next step?'' to improve CIPHER's ability to suggest appropriate actions based on different scenarios.

{(7) Helpful Response Structure}: 
Expert responses in our dataset follow a structure of explanation, reasoning, and examples, aiming to develop the model's ability to provide comprehensive assistance.

{(8) Multiple Turn Conversation}: 
We generate and concatenate multiple-turn conversations to capture the dynamic nature of penetration testing scenarios, creating over 300 long conversations covering entire machine penetration testing sessions.

\begin{figure}[H]

\begin{adjustwidth}{-\extralength}{0cm}
    \centering
    \includegraphics[width=.9\linewidth]{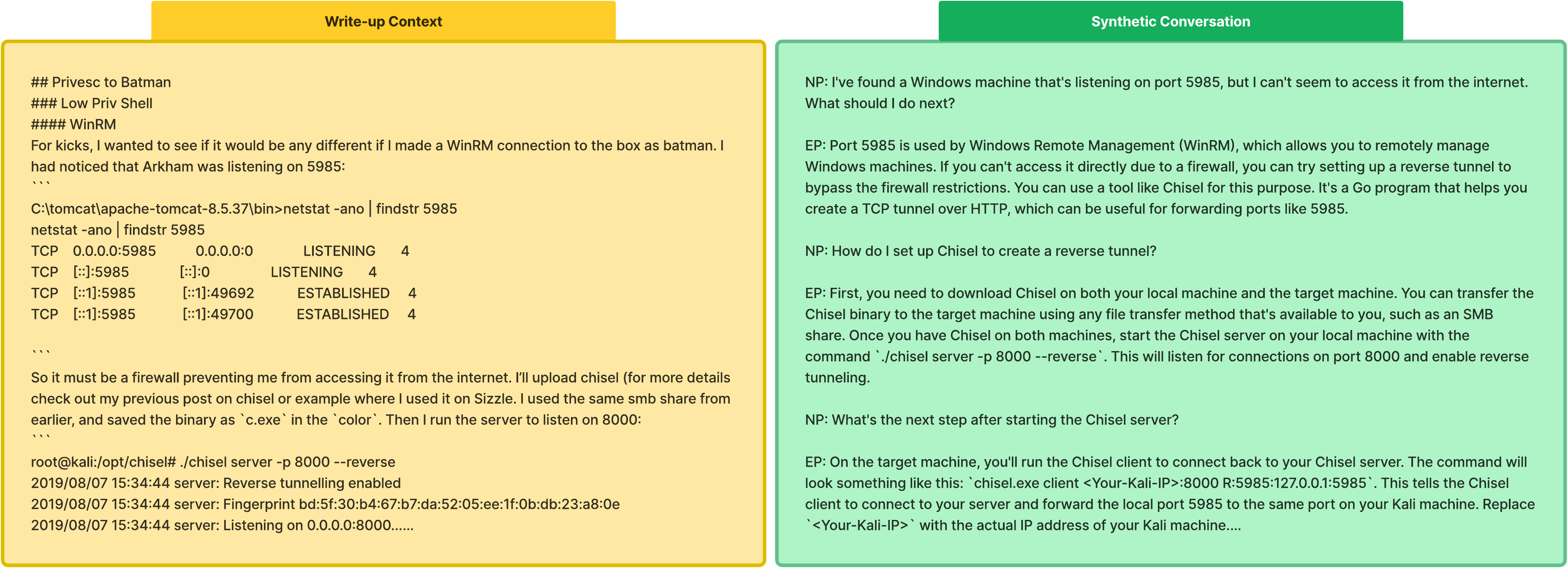}
\end{adjustwidth}
    \caption{\textcolor{black}{{Synthetic conversation} 
 between {Newbie Pentester (NP)} and {Expert Pentester (EP)} generated from real write-up chunk. The expert's experience is reflected in the specific solution, demonstrating knowledge typically gained through practice rather than textbook learning.}}
    \label{fig:conv-gen-result}
\end{figure}

The real dataset sample in Figure \ref{fig:conv-gen-result} demonstrates how our generated conversations effectively capture real-world penetration testing experiences. The expert's response shows depth of knowledge and practical problem-solving, reflecting the expertise typically gained through extensive experience rather than textbook learning. Our conversation conversion method captures the expert problem solving skills such as that related to a WinRM firewall issue, converts it into users problems, 
 then answers with Chisel, which is the preferred tool chosen by the expert.

\subsubsection{\textcolor{black}{Dataset Structure and Pre-Processing}}

\begin{figure}[H]
    \begin{adjustwidth}{-\extralength}{0cm}
    \centering
    \includegraphics[width=.9\linewidth]{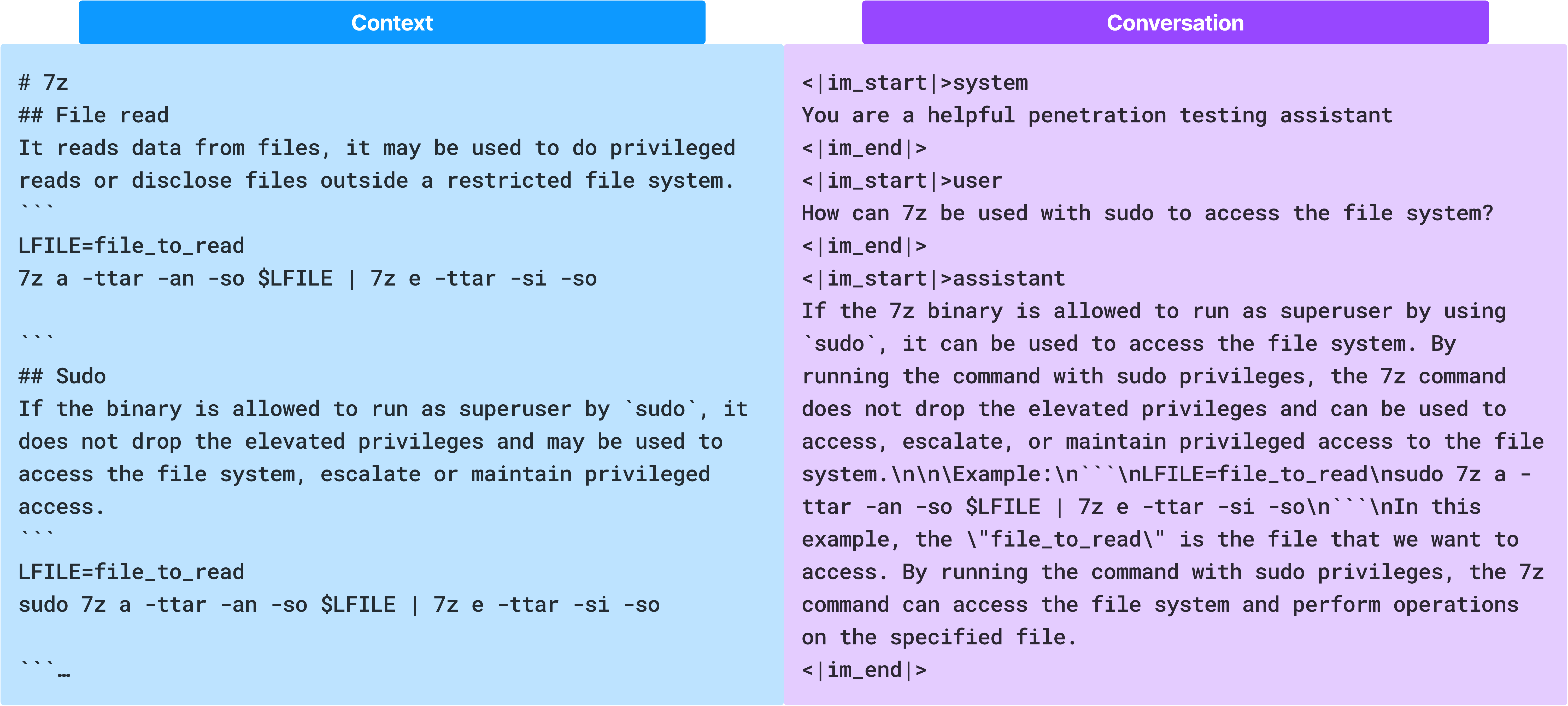}
\end{adjustwidth}
    \caption{\textcolor{black}{{Two} CIPHER training dataset structures: context and conversation type. Note that the '\#' represents header in markdown format while the other is in ChatML format.}}
    \label{fig:dataset-struct} 
\end{figure}

\textcolor{black}{CIPHER trained with two format of datasets. In Figure \ref{fig:dataset-struct}, for the conversation dataset, ChatML format is used, while for base knowledge, the raw context dataset is formatted as markdown with nested header information to prevent loss of information after chunking. All of the context datasets are scraped into markdown files and pre-processed by removing images to reduce hallucinations and token usage cost in training.}

\subsection{Training} \label{section:training}

\begin{figure}[H]
\begin{adjustwidth}{-\extralength}{0cm}
    \centering
    \includegraphics[width=.99\linewidth]{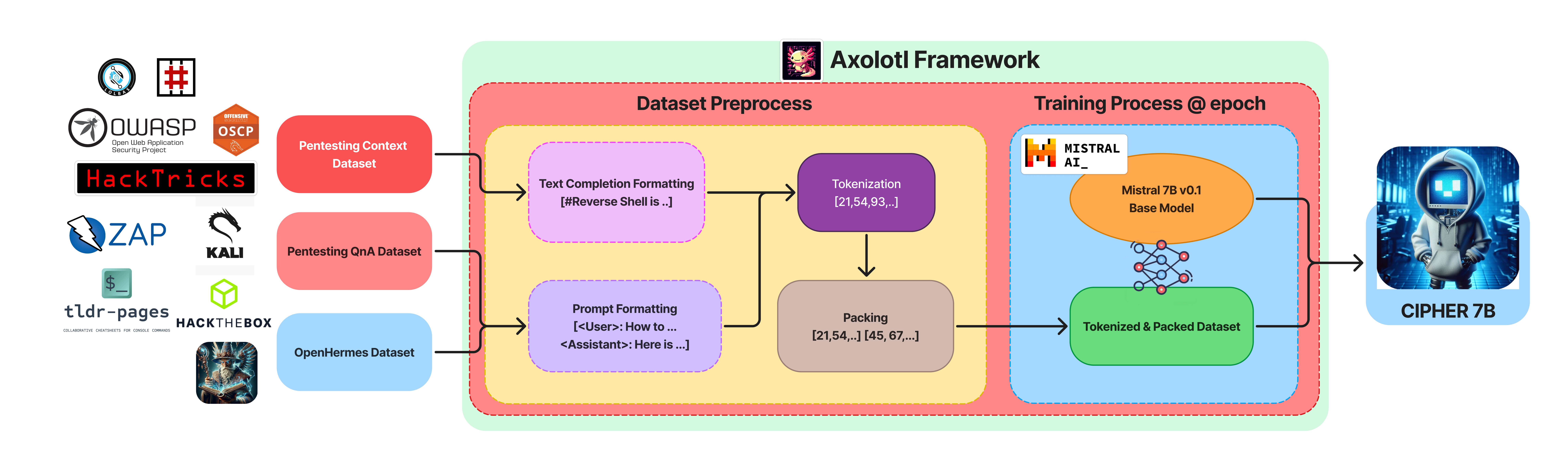}
\end{adjustwidth}
    \caption{CIPHER training pipeline.}
    \label{fig:training-pipeline}
\end{figure}

\textcolor{black}{This section details CIPHER's training process, datasets, and parameters, which leverage the Axolotl framework~\cite{githubGitHubAxolotlaicloudaxolotl} for efficient and easily reproducible training capabilities, as shown in Figure \ref{fig:training-pipeline}.}

\subsubsection{Supervised Fine-Tuning} \label{section:supervised-fine-tuning}
CIPHER's initial training phase uses Mistral v0.3 as a base model combined with the OpenHermes 2.5 dataset with a specialized pentesting dataset. The Axolotl framework manages tokenization and multipacking. We employ a standard text completion approach without specific raw text pretraining data formatting. The OpenHermes 2.5 and pentesting datasets utilize the ChatML format, supporting system, user, and assistant prompts.
The pentesting dataset uses a default system prompt {of} \texttt{"You are a helpful penetration testing assistant."} 
 while the general dataset's prompt remains unmodified. The combined dataset totals approximately 782 million tokens, comprising 500,606,758 tokens of general data and 281,516,236 tokens of supervised data.

\vspace{-9pt}

Fine-tuning is performed on an 8 $\times$ H100 compute cluster. Training the 782 million token dataset, combining pentesting and general data, requires approximately 12 h per epoch. Evaluation reveals optimal loss between epochs 5 and 7, approaching 0.1 loss. We train at a 32 K context length for maximum performance, utilizing a gradient accumulation size of 4 and a micro-batch size of 3. The adamw\_bnb\_8bit optimizer is employed with a cosine scheduler. Full training spans 10 epochs with a $2 \times 10^{-5}$ learning rate without weight decay. Deepspeed is enabled with Zero3 CPU offload optimization.

\subsubsection{Direct Preference Optimization} \label{section:dpo}
Following supervised fine-tuning, we implement Direct Preference Optimization (DPO) to refine model responses using the Argilla DPO Mix 7K~\cite{huggingfaceArgilladpomix7kDatasets} open-source dataset, which is already pre-formatted in ChatML. DPO training is conducted via the Axolotl framework and extends to 4 epochs, surpassing typical epoch counts due to consistently decreasing loss.

\subsection{FARR Augmentation}\label{farr-augmentation}
Currently, evaluating LLMs for penetration testing faces significant challenges:
\begin{itemize}
    \item {No automated benchmarks}: There are no established automatic penetration testing benchmarks for LLMs.
    \item {Reliance on manual testing}: The best available method involves human testers performing manual penetration testing while using an LLM chatbot for guidance.
    \item {Time-consuming}: This manual approach requires approximately 1--4 h per task per evaluator.
    \item {Inconsistent results}: Outcomes can vary significantly based on the evaluator's expertise, condition, and interpretation of the LLM's guidance.
\end{itemize}

These factors make it difficult to efficiently and consistently evaluate LLM performance in penetration testing scenarios, highlighting the need for more standardized and scalable assessment methods in this field.

To address the lack of automated benchmarks for evaluating LLMs in penetration testing, we have developed a novel approach:
\begin{itemize}
    \item {Benchmark Creation}: We have designed a benchmark to measure the accuracy of an LLM's first suggestion in a penetration testing scenario.
    \item {Data Augmentation}: Unlike our previous synthetic conversation data generation method for supervised fine-tuning (SFT), this benchmark augments write-ups into compact, dense lists of information.
    \item {FARR Penetration Testing Flow}: We introduce a new method to augment penetration testing write-ups into a Findings, Action, Reasoning, Result (FARR) Flow. This structure reflects the typical phases of a penetration test, capturing the following:
    Findings: information discovered;
    Action: steps taken based on the findings;
    Reasoning: explanation for the chosen action; and
    Result: outcome of the action.
    \item {Rationale}: We observed that penetration testing write-ups consistently follow this ordered structure, providing a comprehensive view of the testing process and vulnerable points in a system.
\end{itemize}

This approach allows for a more standardized and detailed evaluation of LLM performance in penetration testing scenarios, addressing the limitations of manual evaluation methods.

The Findings, Action, Reasoning, Result (FARR) augmentation process systematically extracts the core vulnerabilities of a target machine from penetration testing write-ups. This structured method produces a FARR Flow, which provides a detailed overview of the penetration testing experience for a specific machine.

A FARR Flow is composed of multiple FARR Steps arranged sequentially. Each FARR Step contains a single set of FARR information, extracted from one specific section of the write-up. As illustrated in Figure \ref{fig:FARR-Flow-Gen-Pipeline}, the FARR Flow Generation Pipeline transforms raw write-ups into a structured format that captures the essence of real-world \mbox{penetration testing}:

\begin{figure}[H]
\begin{adjustwidth}{-\extralength}{0cm}
    \centering
    \includegraphics[width=.95\linewidth]{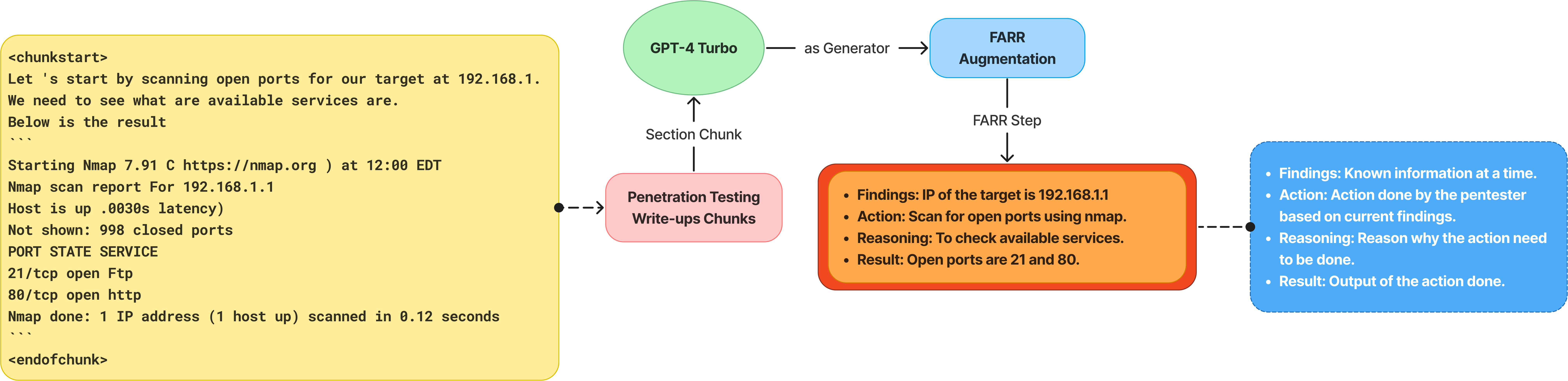}
\end{adjustwidth}
    \caption{\textcolor{black}{Findings
    		, Action, Reasoning, and Result extraction from the real write-ups to construct the real penetration testing experience.}}
    \label{fig:FARR-Flow-Gen-Pipeline}
\end{figure}

The FARR augmentation method offers versatility in evaluating different aspects of a model's knowledge:
\begin{enumerate}
    \item {Findings experience}: Assessing the model's understanding of the required findings for a specific action.
    \item {Reasoning}: Evaluating the model's ability to explain why an action is taken given certain findings.
    \item {Result prediction}: Measuring the model's capability to predict the outcome of \mbox{an action}.
\end{enumerate}

\begin{figure}[H]
\begin{adjustwidth}{-\extralength}{0cm}
    \centering
    \includegraphics[width=.95\linewidth]{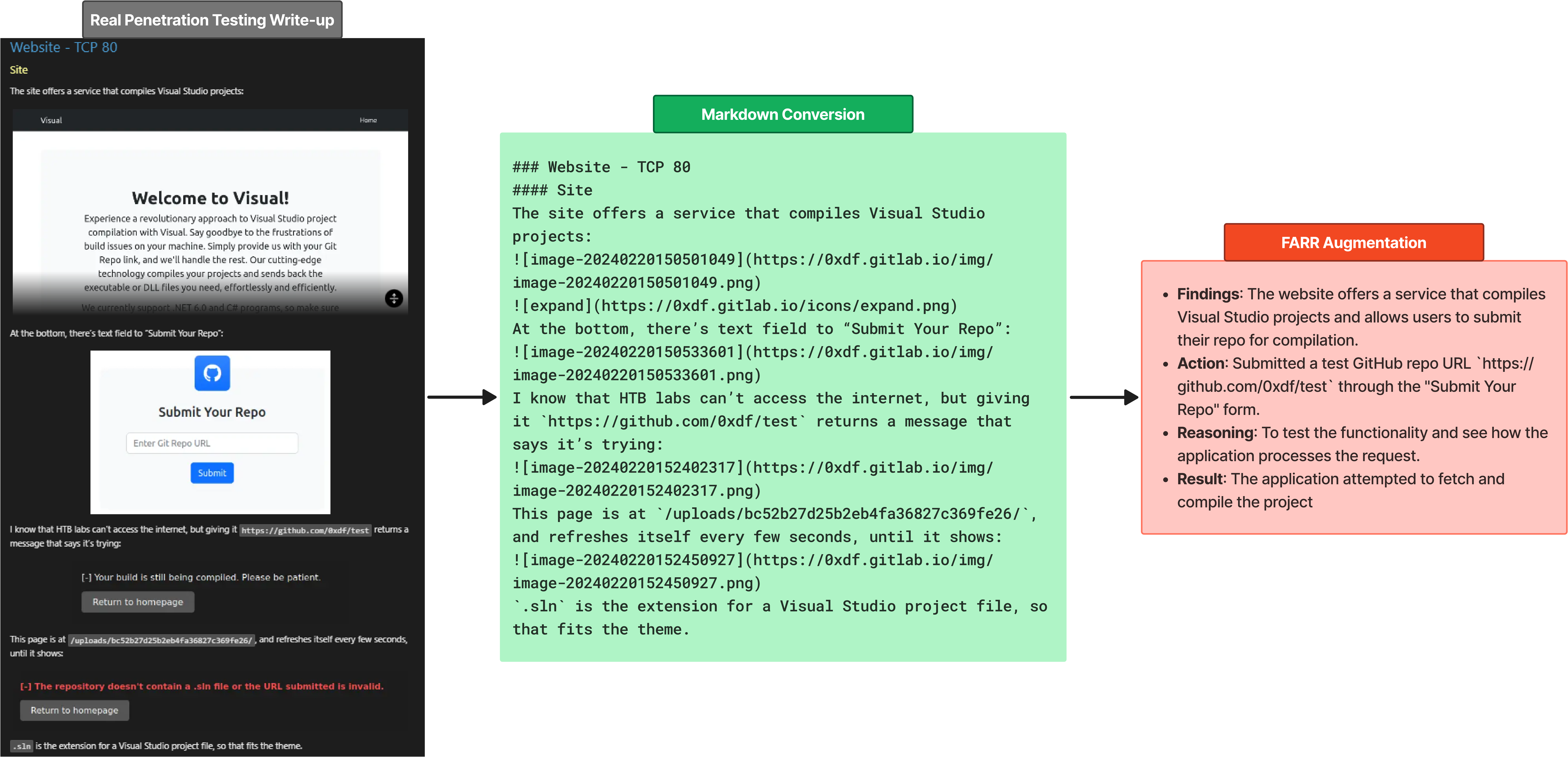}
\end{adjustwidth}
    \caption{\textcolor{black}{Process of FARR augmentation from write-up chunk into compact valuable information, FARR Step. This process is performed until the end of write-ups then combined together as FARR~Flow.}}
    \label{fig:FARR-REAL-AUGMENT}
\end{figure}

\textcolor{black}{Figure \ref{fig:FARR-REAL-AUGMENT} illustrates the augmentation process of transforming write-ups into FARR Steps, where each FARR component contains valuable information. Before the information is augmented, it is scraped and pre-processed into markdown format with nested headers to preserve context when split.}

\textcolor{black}{Some information, such as images will be lost but due to the high quality of the write-up, which is very descriptive, the LLM model can still capture and understand the context clearly. Specifically, it recognizes that the current situation involves testing the input form using a fake URL to evaluate the response.}

\textcolor{black}{Since CIPHER is designed to assist penetration testers with accurate guidance, our evaluation specifically focuses on predicting the reasoning behind actions using known information. This approach allows us to assess how well the model understands and explains decision-making processes in real penetration testing scenarios.
In Section \ref{farr-flow-reasoning-eval}, we use the FARR Flow to evaluate the model's comprehension of real penetration testing experiences, mainly its ability to process gathered information and determine appropriate next steps focused on reasoning. Please note that we do not include FARR Flow augmentation in CIPHER's training dataset to ensure the model’s generalization in providing helpful penetration testing guidance and to allow for a fair comparison with existing models.}

\section{Experiment Results}
In this evaluation, we assess CIPHER's effectiveness in providing accurate guidance for penetration testing. Our assessment consists of several key components:
\begin{itemize}
    \item {General Capabilities}: We use the LLM Eval Harness benchmark to measure CIPHER's general task capabilities, examining how integrating general and penetration testing data affects performance.
    \item {Cybersecurity Expertise}: We evaluate the model's performance on cybersecurity-specific questions to gauge its domain knowledge.
    \item {Pentesting Guidance Accuracy}: Using our novel FARR Flow reasoning evaluation method, we assess CIPHER's ability to guide novice penetration testers through the testing process automatically.
    \item {MITRE ATT\&CK Understanding}: \textls[-25]{We employ the PurpleLlama CyberSecEval~\cite{bhatt_cyberseceval_2024} framework to measure CIPHER's comprehension of the MITRE ATT\&CK \mbox{knowledge base.}}
    \item {Real-world Beginner Use Case Experiment}: To analyze and evaluate the current capabilities of CIPHER in real-world usage assisting a beginner in penetration testing.
\end{itemize}

This multi-faceted approach allows us to comprehensively evaluate CIPHER's capabilities in providing accurate and effective penetration testing guidance.

\subsection{Eval Harness General Capabilities}  
\textcolor{black}{CIPHER's effectiveness in providing penetration testing guidance depends on maintaining coherent conversations. We assess its general language model capabilities using the EleutherAI LLM Eval Harness benchmark~\cite{lintang_sutawika_eleutherailm-evaluation-harness_2023}. This section presents the top two CIPHER models, ranked by their average performance across various tasks in the benchmark.}  

\textcolor{black}{The evaluation is conducted without n-shot prompting, using datasets that measure general-purpose skills (e.g., logical reasoning, commonsense reasoning, world knowledge) and cybersecurity topics. For logical reasoning, critical for context comprehension and step-by-step solutions, we use ARC~\cite{clark2018thinksolvedquestionanswering}, LogiQA~\cite{liu2020logiqachallengedatasetmachine}, and parts of MMLU (Formal Logic and Logical Fallacies)~\cite{hendrycks_measuring_2021}. Commonsense reasoning, essential for practical decision-making and ethical solutions, is evaluated with PiQA~\cite{bisk2019piqareasoningphysicalcommonsense}, SWAG~\cite{zellers2018swaglargescaleadversarialdataset}, and HellaSwag~\cite{zellers2019hellaswagmachinereallyfinish}. For cybersecurity and complex reasoning, we use MMLU's computer science and security sections~\cite{hendrycks_measuring_2021}, along with the Adversarial NLI (ANLI)~\cite{nie2020adversarialnlinewbenchmark}. Lastly, OpenBookQA~\cite{mihaylov2018suitarmorconductelectricity} evaluates common and subject-specific knowledge.  }

{Table \ref{tab:model_performance_logical} presents model performance on logical reasoning tasks, a key factor for LLMs in producing relevant, well-structured output. A lack of proficiency in this area can result in disorganized or out-of-context content. Compared to OpenHermes 2.5, CIPHER shows a slight performance drop of 0.01 in ARC and 0.03 in LogiQA. However, this minimal degradation is acceptable given the large pentesting dataset, indicating no signs of overfitting. CIPHER continues to generalize well across non-pentesting topics. In addition to logical reasoning, common sense reasoning is essential, involving intuitive judgment about feasibility with support from logical reasoning. We also assess complex adversarial examples and world knowledge, which are both important for evaluating deductive reasoning and preventing incorrect conclusions or hallucinations due to limited~knowledge.}

\begin{table}[H]
\caption{Model performance on logical reasoning.}
\newcolumntype{C}{>{\centering\arraybackslash}X}
\begin{tabularx}{\textwidth}{M{5cm}M{2.35cm}M{1.55cm}M{1.cm}M{2cm}}
\toprule
            \textbf{Model} & \textbf{ARC Challenge} & \textbf{ARC Easy} & \textbf{LogiQA} & \textbf{Average Score}\\
        \midrule
        qwen1.5-7b-chat~\cite{Qwen1.57B} & 0.4352 & 0.6831 & 0.2995 & 0.4726\\
        colibri-8b-v0.1~\cite{CyberNativeAIColibri} & 0.4838 & 0.7904 & 0.2565 & 0.5102\\
        lily-cybersecurity-7b-v0.2~\cite{segolilylabsLilyCybersecurity} & 0.4974 & 0.7912 & 0.2903 & 0.5263\\
        meta-llama-3-8b-instruct~\cite{MetaLlama3Instruct} & 0.5265 & 0.8161 & 0.2719 & 0.5381\\
        llama-3-whiterabbitneo-8b-v2.0~\cite{WhiteRabbitNeoLlama3} & 0.5247 & 0.8102 & 0.2873 & 0.5407\\
        openhermes-2.5-mistral-7b~\cite{OpenHermes2.5} & 0.5631 & 0.8329 & 0.2980 & 0.5647\\
        hermes-2-pro-llama-3-8b~\cite{NousResearchHermes} & 0.5520 & 0.8342 & {0.3226} 
 & 0.5696\\
        mistral-7b-instruct-v0.3~\cite{Mistral7BInstruct} & {0.5717} & {0.8418} & 0.3195 & {0.5776}\\
        \textcolor{black}{CIPHER 7B DPO (Ours)} & 0.4838 & 0.7652 & 0.2642 & 0.5044\\
        \textcolor{black}{CIPHER 7B (Ours)} & 0.5503 & 0.8232 & 0.2688 & 0.5475\\
        \bottomrule
\end{tabularx}
\label{tab:model_performance_logical}
\end{table}

\textcolor{black} {In Table \ref{tab:model_performance_common-sense}, CIPHER models are competitive in common sense reasoning benchmarks, trailing the top model in PiQA by only 1.5\%, where OpenHermes 2.5 leads with a score of 0.8156. CIPHER, however, excels in ANLI, with cipher-mistralv0.2-chio-32k-v1.6 achieving the highest score (0.5292), showcasing strong logical reasoning and adversarial handling.}

\textcolor{black}{OpenHermes models lead in both logical and common sense reasoning benchmarks, though CIPHER's lower score in OpenBookQA (0.38 by Hermes-2-Pro-Llama-3-8B) may reflect a trade-off between its penetration testing focus and general knowledge.}

\begin{table}[H]
    \caption{Model performance on common sense reasoning, general logic, and world knowledge.}
 \begin{adjustwidth}{-\extralength}{0cm}
		\newcolumntype{C}{>{\centering\arraybackslash}X}
		\begin{tabularx}{\fulllength}{M{5.8cm}M{1.0cm}M{1.0cm}M{1.3cm}M{2.2cm}M{2.2cm}M{2cm}}
			\toprule
            \textbf{Model} & \textbf{PiQA} & \textbf{Swag} & \textbf{Hellaswag} & \textbf{OpenBook QA} & \textbf{ANLI Average} & \textbf{Average Score} \\
        \midrule
        Lily-Cybersecurity-7B-v0.2~\cite{segolilylabsLilyCybersecurity} & 0.7807 & 0.5610 & 0.6095 & 0.3280 & 0.3955 & 0.5350\\
        Llama-3-Whiterabbitneo-8B-v2.0~\cite{WhiteRabbitNeoLlama3} & 0.7938 & 0.5794 & 0.6093 & 0.3340 & 0.3846 & 0.5402\\
        Meta-Llama-3-8B-Instruct~\cite{MetaLlama3Instruct} & 0.7856 & 0.5700 & 0.5773 & 0.3440 & 0.4653 & 0.5484\\
        Qwen1.5-7b-Chat~\cite{Qwen1.57B} & 0.7546 & 0.5776 & 0.5881 & 0.3220 & 0.5026 & 0.5490\\
        Colibri-8B-v0.1~\cite{CyberNativeAIColibri} & 0.7960 & 0.5866 & 0.5942 & 0.3640 & 0.4676 & 0.5617\\
        Mistral-7B-Instruct-v0.3~\cite{Mistral7BInstruct} & {0.8177} 
& 0.5867 & {0.6475} & 0.3540 & 0.4560 & 0.5724\\
        Hermes-2-Pro-Llama-3-8B~\cite{NousResearchHermes} & 0.8003 & {0.6026} & 0.6266 & {0.3800} & 0.4776 & 0.5774\\
        OpenHermes-2.5-Mistral-7B~\cite{OpenHermes2.5} & 0.8156 & 0.5935 & 0.6310 & 0.3460 & 0.5116 & {0.5795}\\
        \textcolor{black}{CIPHER 7B DPO (Ours)} & 0.7753 & 0.5682 & 0.5922 & 0.2920 & 0.4903 & 0.5436\\
        \textcolor{black}{CIPHER 7B (Ours)} & 0.7965 & 0.5935 & 0.6288 & 0.3260 & {0.5194} & 0.5728\\
      \bottomrule
		\end{tabularx}
    \label{tab:model_performance_common-sense}
\end{adjustwidth}
\end{table}

\textcolor{black}{In summary, CIPHER models excel in logical reasoning but show a slight trade-off in general knowledge, likely due to their specialization. Future efforts should aim to balance domain-specific expertise with broader knowledge, with continued evaluation on specialized topics like computer science and security.}

\textcolor{black}{Table \ref{tab:model_performance_mmlu} shows the results for specialized MMLU topics in cybersecurity. Despite being trained on penetration testing datasets, CIPHER underperforms compared to base models on general tasks, suggesting that while optimized for penetration testing, it may be less effective on broader MMLU tasks.}

\textcolor{black}{Qwen1.5-7B-Chat excels in high school and college computer science, while Llama-3-WhiteRabbitNeo-8B-v2.0 leads in computer security. CIPHER, though not the top performer, is consistent across categories, blending general competency with specialized penetration testing knowledge, meeting LLM assistant standards.}

\begin{table}[H]
    \caption{Specialized MMLU for cyber security-related topics.}
\begin{adjustwidth}{-\extralength}{0cm}
    	\newcolumntype{C}{>{\centering\arraybackslash}X}
		\begin{tabularx}{\fulllength}{M{5.8cm}M{2cm}M{1.5cm}M{1.5cm}M{1.5cm}M{1.5cm}M{1.5cm}}
			\toprule
        \textbf{Model} & \textbf{High School CompSci} & \textbf{Computer Security} & \textbf{College CompSci} & \textbf{Formal Logic} & \textbf{Logical Fallacies} & \textbf{Average Score}\\
        \midrule
        colibri-8B-v0.1~\cite{CyberNativeAIColibri} & 0.5900 & 0.6800 & 0.4200 & 0.4127 & 0.7055 & 0.5616\\
        lily-cybersecurity-7B-v0.2~\cite{segolilylabsLilyCybersecurity} & 0.5800 & 0.6800 & 0.5600 & 0.3333 & 0.7055 & 0.5718\\
        openhermes-2.5-Mistral-7b~\cite{OpenHermes2.5} & 0.7100 & 0.7200 & 0.4500 & 0.4048 & 0.7607 & 0.6091\\
        mistral-7B-Instruct-v0.3~\cite{Mistral7BInstruct} & 0.6200 & 0.6700 & 0.5400 & 0.4524 & {0.7730} & 0.6111\\
        hermes-2-pro-llama-3-8B~\cite{NousResearchHermes} & 0.6700 & 0.7100 & 0.4700 & 0.4762 & 0.7669 & 0.6186\\
        qwen1.5-7B-Chat~\cite{Qwen1.57B} & {0.7200} & 0.7600 & {0.6100} & 0.4127 & 0.6748 & 0.6355\\
        llama-3-whiterabbitneo-8B-v2.0~\cite{WhiteRabbitNeoLlama3} & 0.6700 & {0.7800} & 0.5400 & 0.4762 & 0.7239 & 0.6380\\
        meta-llama-3-8B-Instruct~\cite{MetaLlama3Instruct} & 0.6800 & 0.7300 & 0.5200 & {0.5000} & 0.7607 & {0.6381}\\
       \textcolor{black}{ CIPHER 7B DPO (Ours)} & 0.6600 & 0.6800 & 0.4800 & 0.3651 & 0.6933 & 0.5757\\
        \textcolor{black}{CIPHER 7B (Ours)} & 0.6100 & 0.7300 & 0.5500 & 0.3651 & 0.7301 & 0.5970\\
         \bottomrule
		\end{tabularx}
    \label{tab:model_performance_mmlu}
\end{adjustwidth}
\end{table}

\subsection{Cybersecurity Knowledge Evaluation}
In this section, we conduct a two-part evaluation of CIPHER:
\begin{itemize}
    \item Pentesting QA Evaluation: CIPHER is developed to bridge the gap between novice and expert. Therefore, the capability to explain 
     penetration testing knowledge is very crucial. We utilize the pentesting evaluation topics from the \texttt{preemware/pentesting-eval} Huggingface repository~\cite{preemware_preemwarepentesting-eval_2024}. This dataset consists of question and explanation pairs from GPT-4. We use the questions as prompts and the explanations as ground truth. Model responses are then compared to the ground truth and assessed by the Qwen1.5 72B Chat model to produce a score indicating alignment with the \mbox{ground truth}.
    \item Comparative Analysis: We compare the results from the LM evaluation harness to identify any performance degradation in general tasks after training various CIPHER models with pentesting datasets.
\end{itemize}

\textcolor{black}{As shown in Table \ref{tab:Pentest evaluation Qwen}, CIPHER models achieve the highest average score of around 83\%, outperforming other LLMs in pentesting-specific evaluations. While models like OpenHermes 2.5 excel in general MMLU tasks, they fall short in comparison to CIPHER on specialized pentesting explanation tasks. Similarly, models like Lily Cybersecurity and WhiteRabbitNeo perform poorly, likely due to overfitting or a decline in \mbox{explanatory capabilities}.}

\begin{table}[H]
    \caption{Pentesting QA evaluation.}
    \newcolumntype{C}{>{\centering\arraybackslash}X}
\begin{tabularx}{\textwidth}{p{9cm}c}
\toprule
            \textbf{Model} & \textbf{Average} \\
        \midrule
        lily-cybersecurity-7B-v0.2~\cite{segolilylabsLilyCybersecurity} & 0.5688 \\
        llama-3-whiterabbitneo-8B-v2.0~\cite{WhiteRabbitNeoLlama3} & 0.5759 \\
        colibri-8b-v0.1~\cite{CyberNativeAIColibri} & 0.5887 \\
        hermes-2-pro-llama-3-8B~\cite{NousResearchHermes} & 0.6626 \\
        qwen1.5-7B-chat~\cite{Qwen1.57B} & 0.6846 \\
        meta-llama-3-8B-instruct~\cite{MetaLlama3Instruct} & 0.7008 \\
        openhermes-2.5-Mistral-7B~\cite{OpenHermes2.5} & 0.7157 \\
        mistral-7B-Instruct-v0.3~\cite{Mistral7BInstruct} & 0.7896 \\
        \textcolor{black}{CIPHER 7B (Ours)} & {0.8302} 
 \\
        \textcolor{black}{CIPHER DPO 7B (Ours)} & {0.8394} 
 \\
        \bottomrule
\end{tabularx}
    \label{tab:Pentest evaluation Qwen}
\end{table}

\textcolor{black}{No other cybersecurity model approaches CIPHER's performance. The substantial gap highlights CIPHER’s effectiveness in pentesting tasks, despite trade-offs in general knowledge areas seen in previous benchmarks.}

\begin{figure}[H]
        \includegraphics[scale=0.5]{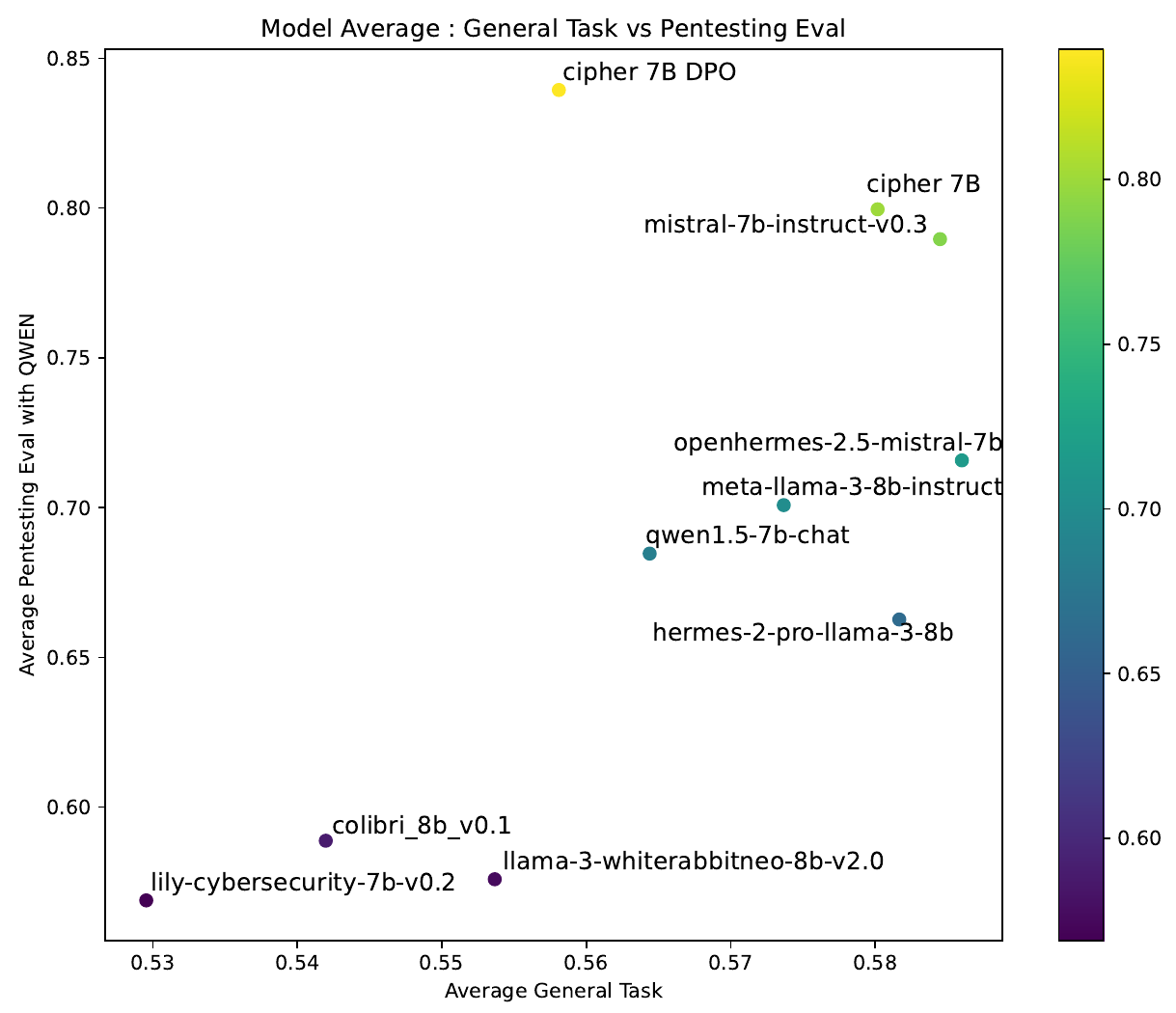}
\caption{{Comparison of models on General Task and Pentesting Evaluation:} The plot illustrates the performance of various models, comparing their average scores. Higher placement on both axes indicates better performance across both domains.}
\label{fig:average_top_cipher_model} 
\end{figure}

\textcolor{black}{The general task performance (Figure \ref{fig:average_top_cipher_model}) shows CIPHER excelling in pentesting tasks while maintaining competitive general task performance. It outperforms cybersecurity-focused models like WhiteRabbitNeo and Lily, which show the lowest results in both general and pentesting-specific tasks. This highlights CIPHER’s effective training approach, enhancing pentesting capabilities without sacrificing general knowledge.}

\textcolor{black}{The CIPHER DPO model performs best in pentesting but scores lower on general tasks, indicating possible overfitting toward specialized knowledge. This trade-off suggests that further optimization could balance general and pentesting performance more effectively. Despite this, CIPHER consistently outperforms other models in both domains, especially in pentesting, positioning it as a strong pentesting assistant}.

\textcolor{black}{In summary, CIPHER non-DPO offers a good balance between general and specialized knowledge, while the DPO version excels in pentesting but requires improvements in general task performance.}

\subsection{Penetration Testing Guidance Accuracy Evaluation (FARR Flow Reasoning)} \label{farr-flow-reasoning-eval}
The general and cybersecurity knowledge question benchmarks alone are not enough to measure how good the model is in guiding the user in the penetration testing process. We formulated a realistic situation combining much information and a dynamic situation obtained from a 
real penetration testing process. 

\begin{figure}[H]

\begin{adjustwidth}{-\extralength}{0cm}
\centering 
    \includegraphics[width=1\linewidth]{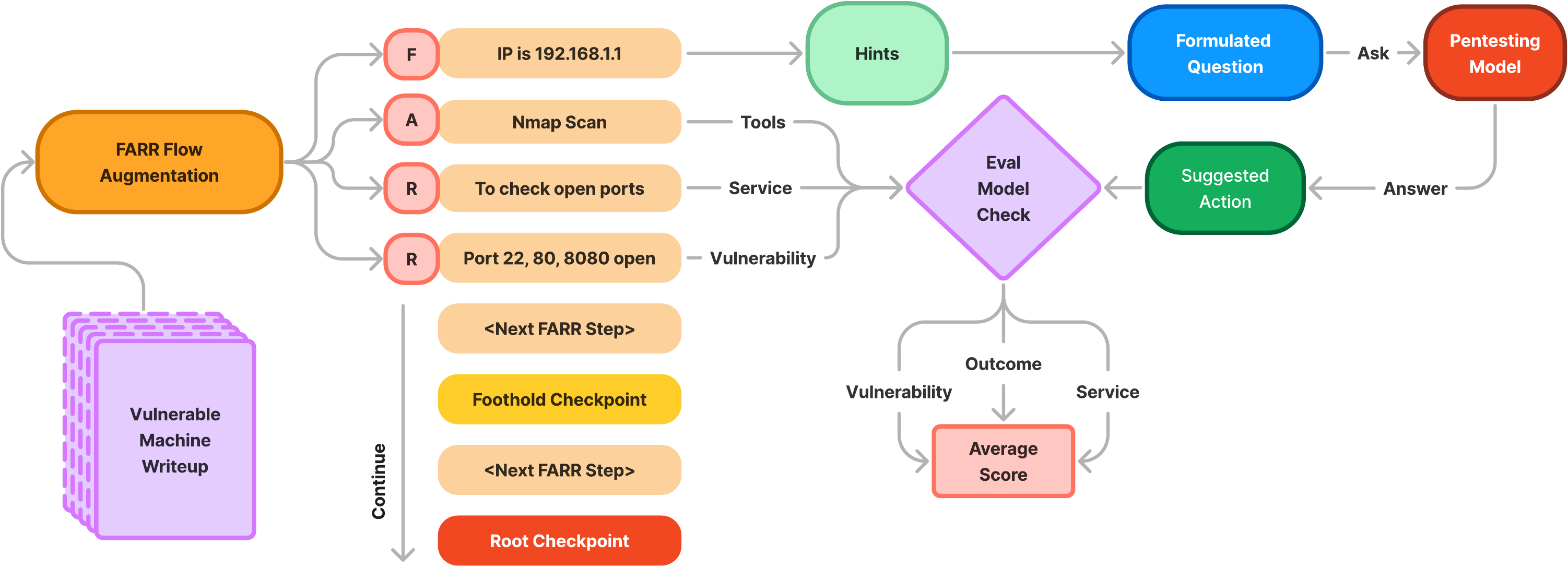}
\end{adjustwidth}
\caption{FARR Flow evaluation.}
\label{fig:FARR-Flow-Evaluation}
\end{figure}

In Section \ref{farr-augmentation}, after the augmentation process and obtaining the full flow of the machine penetration testing steps augmented as FARR Flow, we constructed a guidance reasoning test for the model by asking it what to do next when we have only the Findings. FARR Flow consists of multiple sets of Findings, Actions, Reasonings, and Results, as shown in Figure \ref{fig:FARR-Flow-Evaluation}, where F is Finding, A is Action, R is Reasoning, and the last R is Result.

Figure ~\ref{fig:FARR-Flow-Evaluation} also illustrates how we evaluate the model using FARR Flow. At each step, we utilize the Findings as known information. We formulate questions for the model to predict the most likely action to do based on the current hint, which is constructed from the accumulated previous and current findings. The model's suggested response is then evaluated against Llama 3 70B to assess the relevance of the answer across three aspects: whether the model's answer anticipates the same outcome, targets the same service, and focuses on the same vulnerability type as the reference. Each criterion is rated with up to \mbox{three points} for relevance. Using these three aspects, we measure the average performance of each model on each Hack The Box machine in the FARR Flow evaluation.

Our guidance reasoning test using the FARR Flow evaluation algorithm is shown in Algorithm \ref{alg:farrstepquestion}. At the first question, the user does not know about the target machine, often only the IP information or open ports. Therefore, we only provide the current findings. However, regardless of the suggested answer to the following question, we will still provide the model with the result of the previous findings to ensure the model understands the current situation. Section \ref{improvebenchmark} discusses the possibility of improvement in this \mbox{question formulation}.

\begin{algorithm}[H]
\caption{Formulating FARR Step Question}
\label{alg:farrstepquestion}

\textbf{Input:} JSON file containing a list of findings, actions, reasonings, and results (\texttt{Machine\_X\_FARR\_flow.json}) \\
\textbf{Output:} Suggested next action from inference model

\begin{algorithmic}[H]
\STATE Load \texttt{Machine\_X\_FARR\_flow.json}
\STATE Parse JSON content into \texttt{FARRflow} list of dictionaries
\STATE Initialize \texttt{all\_findings} as an empty string

\FOR{each \texttt{flow} in \texttt{FARRflow}}
    \STATE \texttt{current\_findings} $\leftarrow$ \texttt{flow[``Findings'']}
    \STATE \texttt{current\_result} $\leftarrow$ \texttt{flow[``Result'']}

    \STATE \texttt{question\_prompt} $\leftarrow$ ``Below are my current findings:\\\texttt{all\_findings}\\\texttt{current\_findings}\\What is the most likely action to do next?
    Answer with one specific action only, not more than that.''
    
    \STATE \texttt{model\_output} $\leftarrow$ \texttt{inference\_model(question\_prompt)}

    \STATE \texttt{all\_findings} $\leftarrow$ \texttt{all\_findings} + ``\\\texttt{current\_findings}, \texttt{current\_result}''
\ENDFOR

\STATE Return \texttt{model\_output} as the model suggested action

\end{algorithmic}
\end{algorithm}

\begin{figure}[H] 
\begin{adjustwidth}{-\extralength}{0cm}
    \centering
    \begin{subfigure}[b]{0.45\textwidth}
        \centering
        \includegraphics[scale=0.5]{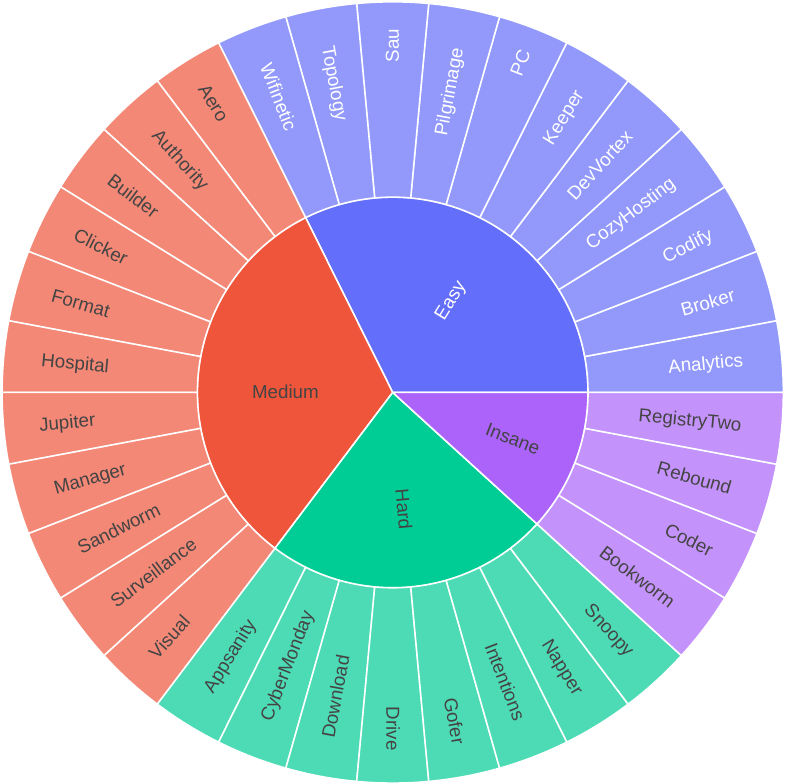}
        \caption{}
        \label{fig:htb-challenges}
    \end{subfigure}
    \hspace{0.05\textwidth}
    \begin{subfigure}[b]{0.45\textwidth}
        \centering
        \includegraphics[scale=0.5]{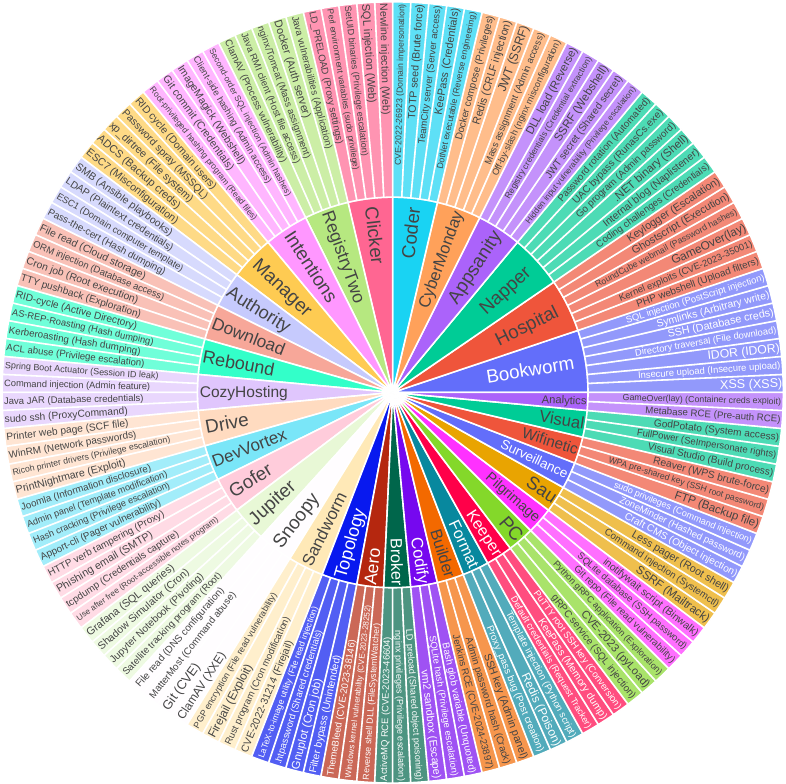}
        \caption{}
        \label{fig:machine-sunburst}
    \end{subfigure}
\end{adjustwidth}
    \caption{{Hack }The Box machines: (\textbf{a}) Hack The Box machines selected for challenges sorted by difficulties, (\textbf{b}) Hack The Box machines selected for evaluation and corresponding attack \mbox{vector scope}.}
    \label{fig:HackTheBox}
\end{figure}

CIPHER training data use the 0xdf write-up~\cite{0xdf0xdfHacks} dataset until early September 2023. We took whole machine write-up samples from mid-September 2023 until the FARR Flow benchmark development started in early May 2024 without cherry-picking. We only excluded incomplete machine write-ups and separated them, resulting in 35 machine write-ups augmented as FARR Flow for the reasoning test, with different difficulties as shown in Figure \ref{fig:HackTheBox}a. Note that these machines are completely unknown to CIPHER. Overall, there are 35 Hack The Box machines with 136 vulnerability topics, all in 2124 
 formulated~questions.
 
Figure \ref{fig:HackTheBox}b presents different attack vectors in this evaluation. The scope includes diverse vulnerabilities and techniques used to penetrate the machine. Each machine has a different attack vector to achieve a foothold from easy to insane level. Various privilege escalation techniques are needed to achieve better scores, and CVE knowledge is necessary to access some machines. This covers better coverage and penetration testing dynamics than traditional question-and-answer~evaluation.

\begin{figure}[H]

\begin{adjustwidth}{-\extralength}{0cm}
    \centering
    \includegraphics[width=.98\linewidth]{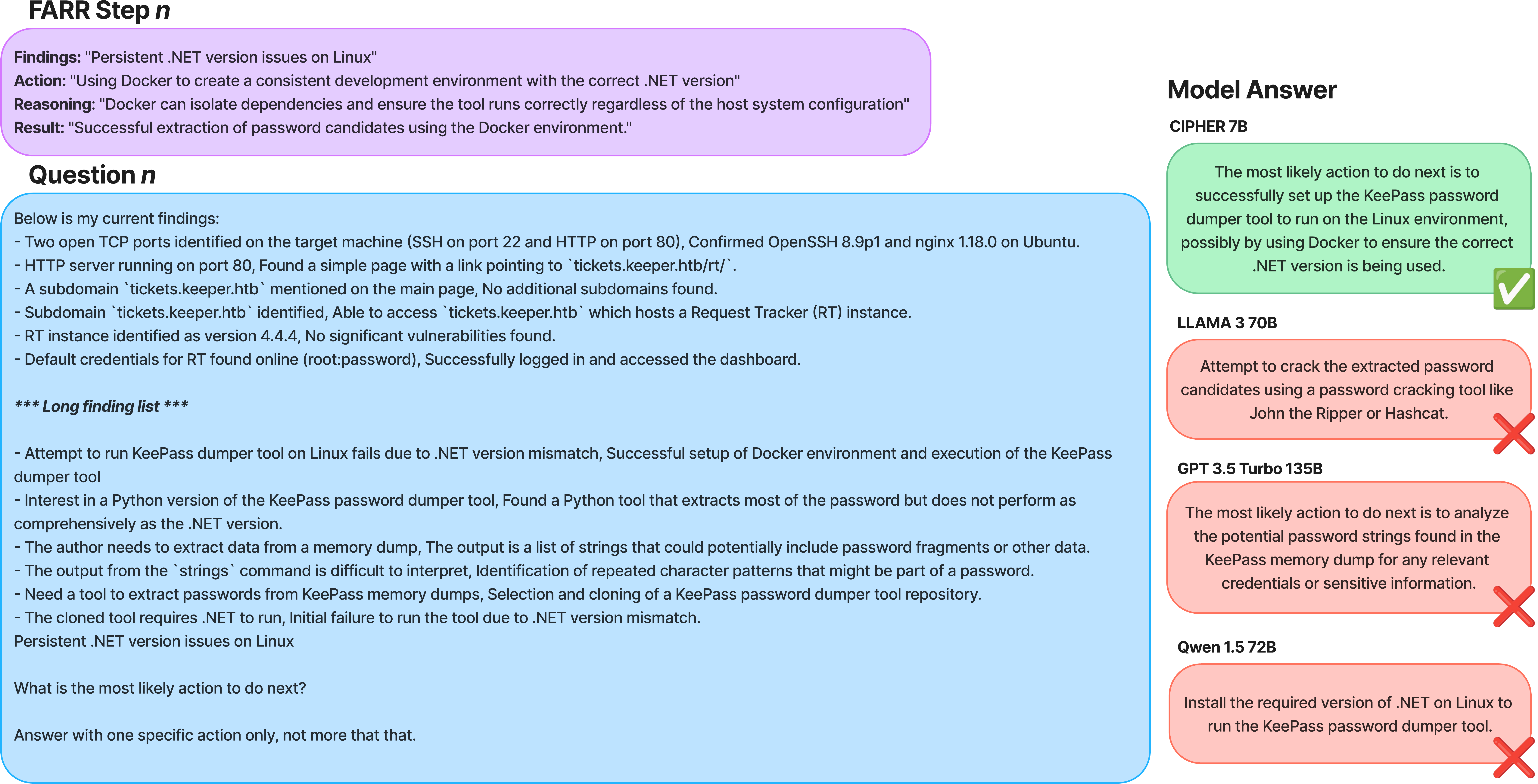}
\end{adjustwidth}
\caption{\textcolor{black}{{Real FARR} 
 Flow augmented question from Hack The Box Keeper machine}. Notes (***: the rest of middle part findings.)}
\label{fig:FARR-Flow-Question}
\end{figure}

\textcolor{black}{Figure \ref{fig:FARR-Flow-Question} shows real example of a question augmented from FARR Flow from a Hack The Box Keeper machine, where it contains a list of findings and results to provide a situation of the user's current penetration testing progress.} Based on the question, the last situation is the current findings in the current FARR step; the model needs to understand the current progress first and then suggest the most likely action to be taken next. We ask for the most likely action to extract the best action suggested from the model to measure the accuracy of the vulnerability points, the service targeted, and the expected outcome from the~action.

\textcolor{black}{The prompt question in Figure \ref{fig:FARR-Flow-Question} provides valuable information about the current status of the penetration testing progress, including open ports, service versions, subdomains, running instances, credentials, and both failed and successful attempts. These details are typically noted by penetration testers. For beginners in penetration testing, however, this information can be overwhelming, making it difficult to determine the appropriate next steps. In real-world scenarios, beginners often turn to forums to ask experts or more experienced penetration testers for hints, or what is commonly referred to as a ``nudge'' in the~community.}

\textcolor{black}{This is where CIPHER bridges the gap. According to the write-ups augmented in FARR Step {Figure} 
 \ref{fig:FARR-Flow-Question}, the current finding is ``\textit{Persistent .NET version issues on Linux},'' and the correct action is to ``\textit{Use Docker to create a consistent development environment with the correct .NET version}.'' However, all three of the other models failed to suggest this. Llama 3 70B, for instance, suggested attempting to extract a password, which is irrelevant. GPT-3.5 Turbo proposed analyzing a memory dump, which does not make sense since it was already stated that the tool could not run. Qwen 1.5 72B suggested installing .NET on Linux, which is also incorrect because the question clearly states that there is a persistent issue with .NET on Linux. CIPHER, however, responded correctly and aligned with the reference by suggesting that the user run the tool in Docker with the correct .NET tools. When provided with complex information from the penetration testing process, CIPHER was able to accurately determine the next step, outperforming even models with ten times the number of parameters, such as Llama 3 70B, GPT-3.5 Turbo, and Qwen 1.5 72B.}

\textcolor{black}{The most relevant models are selected for this evaluation, including available small pentesting or cybersecurity models, and we also included the best open-source state-of-the-art (SOTA) model as the judge: Llama 3 70B. To ensure our benchmark is compatible with larger parameter models, we included the Llama 3 70B and Qwen 1.5 72B Chat scores in our benchmark. This allows us to assess how well our model's guidance compares to larger models out of CIPHER's size.}

\subsubsection{Penetration Testing Guidance Accuracy Performance}
Table \ref{tab:model_performance_detailed} shows the three criteria measured in our FARR Flow evaluation for each model tested. Our model CIPHER achieves the highest average score compared to other penetration testing models and large SOTA general models like Llama 3 70b and Qwen1.5 72B. Llama 3 70B scores slightly higher (by 0.19) in outcome prediction, but CIPHER guides better in service aimed and vulnerability targeted, with a 0.52 higher score overall at 52.59. However, compared to other penetration testing models similar to CIPHER's size, there is a significant increase in performance, as big as 9.75\% from Colibri 0.1 8B.

Since the judge LLM used was Llama 3 70B, there might be some bias in the result causing Llama 3 to score higher on its answer. Conversely, Qwen1.5 72B achieves a score similar to GPT3.5, which seems more accurate and free from judgment bias.

\begin{table}[H]
        \caption{Model performance on FARR Flow reasoning test.}
    \newcolumntype{C}{>{\centering\arraybackslash}X}
\begin{tabularx}{\textwidth}{p{5cm}cccc}
\toprule
            \textbf{Model} & \textbf{Outcome} & \textbf{Service} & \textbf{Vulnerability} & \textbf{Total Avg} \\
        \midrule
        whiterabbitneo-7b~\cite{WhiteRabbitNeoLlama3} & 25.81 & 38.78 & 27.49 & 30.69 \\
        lily-7b~\cite{segolilylabsLilyCybersecurity} & 29.35 & 43.77 & 35.57 & 36.23 \\
        qwen-7b~\cite{Qwen1.57B} & 36.50 & 51.18 & 40.11 & 42.60 \\
        colibri-0.1-8b~\cite{CyberNativeAIColibri} & 40.85 & 56.61 & 46.30 & 47.92 \\
        hermes2pro-llama3-8b~\cite{NousResearchHermes} & 41.27 & 58.13 & 47.15 & 48.85 \\
        mistral-ins-0.2-7b~\cite{Mistral7BInstruct} & 41.36 & 59.91 & 46.60 & 49.29 \\
        qwen-72b~\cite{Qwen1.572B} & 43.57 & 59.05 & 46.84 & 49.82 \\
        gpt-3.5-turbo~\cite{ouyang2022training} & 43.57 & 59.31 & 47.45 & 50.11 \\
        llama-3-70b-ins~\cite{MetaLlama3Instruct} & {44.92} 
 & 61.67 & 49.62 & 52.07 \\
        \textcolor{black}{CIPHER DPO 7B (Ours)} & {44.73} 
 & {62.65} & {50.39} & {52.59} \\
        \bottomrule
\end{tabularx}
\label{tab:model_performance_detailed}
\end{table}

Smaller models with around 7B parameters score below 49.30, led by Mistral Instruct v0.2 7B, which excels at general instruction. Hermes2pro-llama3-8b, known for its agentic and tool usage prowess, still cannot achieve good guidance on penetration testing. Colibri 7B achieves 47.92 overall; despite its strength in QnA evaluation, it fails to correctly guide 50 percent of the questions, except for aiming at the correct service (56.61).

Qwen 1.5 7B only achieves 42.60, often providing defensive responses instead of offensive ones. The performance gap between Qwen1.5 72B and 7B suggests that model size scaling can affect the model's benign level. Based on our experiments, Qwen 7B tends to give defensive solutions when asked about the next steps.

Other open-source penetration testing cybersecurity models perform poorly. Lily 7B, based on Mistral Instruct fine-tuning, provides poor guidance at 36.23, followed by Whiterabbitneo 7B at 30.69. Whiterabbitneo's poor performance is due to its bias towards answering everything with code and often producing incoherent responses.

\textcolor{black}{In summary, most penetration testing models struggle to handle the large volume of information and often suggest inaccurate next steps. However, state-of-the-art models like Llama 3 (70B) achieve the highest scores in outcome prediction. Despite this, our CIPHER DPO (7B) model outperforms even the largest models in terms of suggestion accuracy. This means that CIPHER guides users more effectively, enabling them to penetrate vulnerable machines faster, achieve a foothold, and gain the highest privilege accounts more efficiently than models ten times its size.}

\subsubsection{Performance by Machine Difficulties}

To evaluate the model's performance across different complexity levels, we clustered the scores based on machine difficulties. This allows us to identify the areas where the model performs better or worse depending on the complexity. Table \ref{tab:model_performance_difficulty} shows that our model CIPHER performs better on average due to its distinguished high scores on easy and insane machines, with 56.25 and 50.96 points, respectively, creating the highest gaps of about 1.69 and 2.22 points from the second-place model. Llama 3 70B takes second place, with slightly better scores on medium and hard machines by 0.52 and 0.43 points. This demonstrates our model's strength in penetration testing reasoning guidance compared to other penetration testing models, even those with ten times the parameter size.

\begin{figure}[H]
\begin{adjustwidth}{-\extralength}{0cm}
\centering
\includegraphics[width=0.95\linewidth]{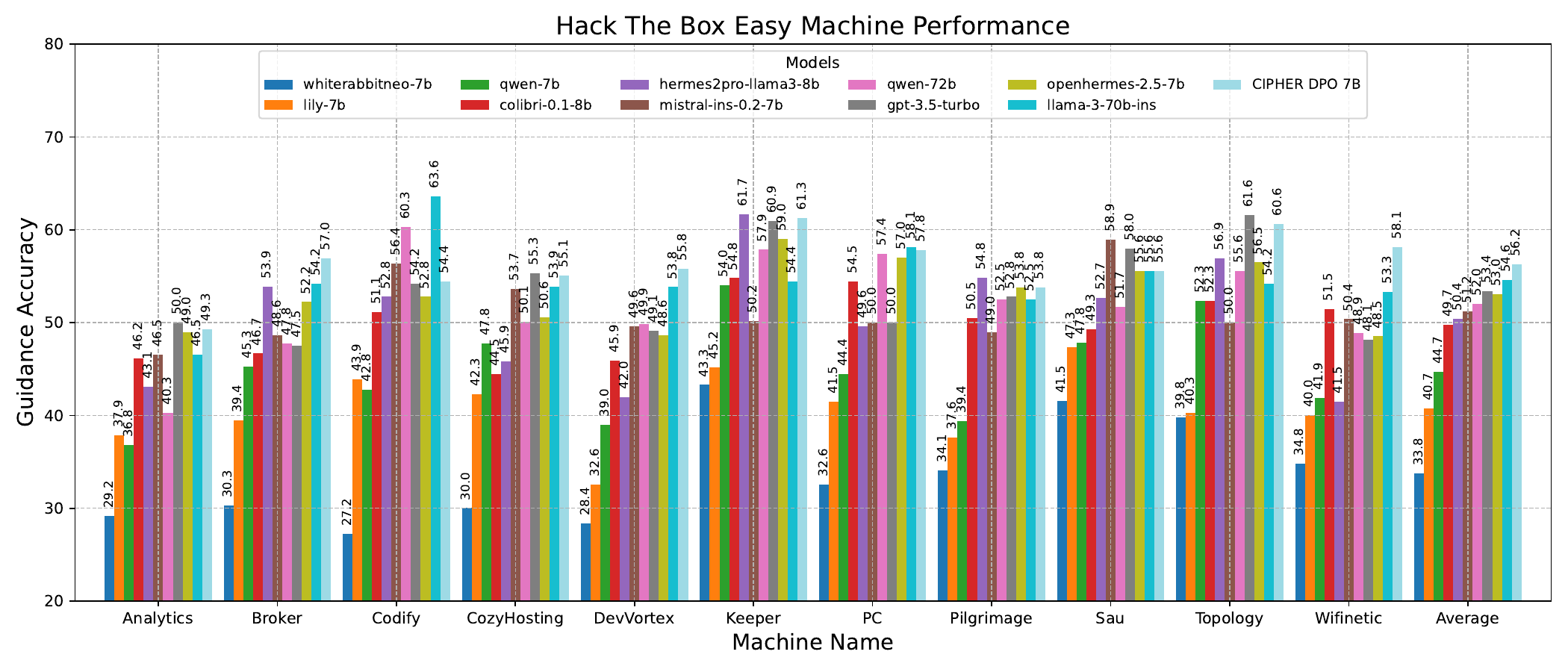}
\end{adjustwidth}
\caption{\textcolor{black}{Easy machine performance results}.}
\label{fig:easy}
\end{figure}

\textcolor{black}{In the detailed results for easy-difficulty machines shown in Figure \ref{fig:easy}, the accuracy scores of various models on easy-difficulty Hack The Box vulnerable machines are compared. CIPHER dominates, outperforming Llama 3 70B with an accuracy of 56.25. This is attributed to CIPHER's extensive training with vulnerable machines in its dataset, which enhances its ability to generalize in these easier challenges, typically involving single-step problem-solving or a single vulnerability.}

\begin{table}[H]
    \caption{Model performance FARR Flow reasoning test sorted by machine difficulty scores.}
          \newcolumntype{C}{>{\centering\arraybackslash}X}
\begin{tabularx}{\textwidth}{p{5cm}ccccc}
\toprule
            \textbf{Model} & \textbf{Easy} & \textbf{Medium} & \textbf{Hard} & \textbf{Insane} & \textbf{Average} \\
        \midrule
        whiterabbitneo-7b~\cite{WhiteRabbitNeoLlama3} & 33.75 & 30.01 & 28.47 & 28.75 & 30.25 \\
        lily-7b~\cite{segolilylabsLilyCybersecurity} & 40.73 & 34.66 & 34.50 & 32.03 & 35.48 \\
        qwen1.5-7b {\cite{Qwen1.57B}} & 44.68 & 41.31 & 42.02 & 41.88 & 42.47 \\ 
        colibri-0.1-8b~\cite{CyberNativeAIColibri} & 49.74 & 47.38 & 45.82 & 48.75 & 47.92 \\
        hermes2pro-llama3-8b~\cite{NousResearchHermes} & 50.43 & 47.76 & 48.54 & 48.37 & 48.78 \\
        mistral-ins-0.2-7b~\cite{Mistral7BInstruct} & 51.21 & 49.34 & 47.27 & 47.89 & 48.93 \\
        qwen1.5-72b~\cite{Qwen1.572B} & 52.02 & 47.81 & 50.88 & 47.67 & 49.60 \\
        gpt-3.5-turbo~\cite{ouyang2022training} & 53.41 & 48.18 & 49.67 & 47.72 & 49.74 \\
        openhermes-2.5-7b~\cite{OpenHermes2.5} & 53.05 & 48.51 & 50.27 & 48.50 & 50.08 \\
        llama-3-70b-ins~\cite{MetaLlama3Instruct} & {54.56} 
 & {51.28} 
 & {51.48} & {48.74} & 51.52 \\
        \textcolor{black}{{CIPHER DPO 7B (Ours)}} & {56.25} & {50.80} & {51.05} & {50.96} & {52.26} \\
        \bottomrule
\end{tabularx}
    \label{tab:model_performance_difficulty}
\end{table}
\vspace{-9pt}

\begin{figure}[H]
\begin{adjustwidth}{-\extralength}{0cm}
\centering
\includegraphics[width=0.95\linewidth]{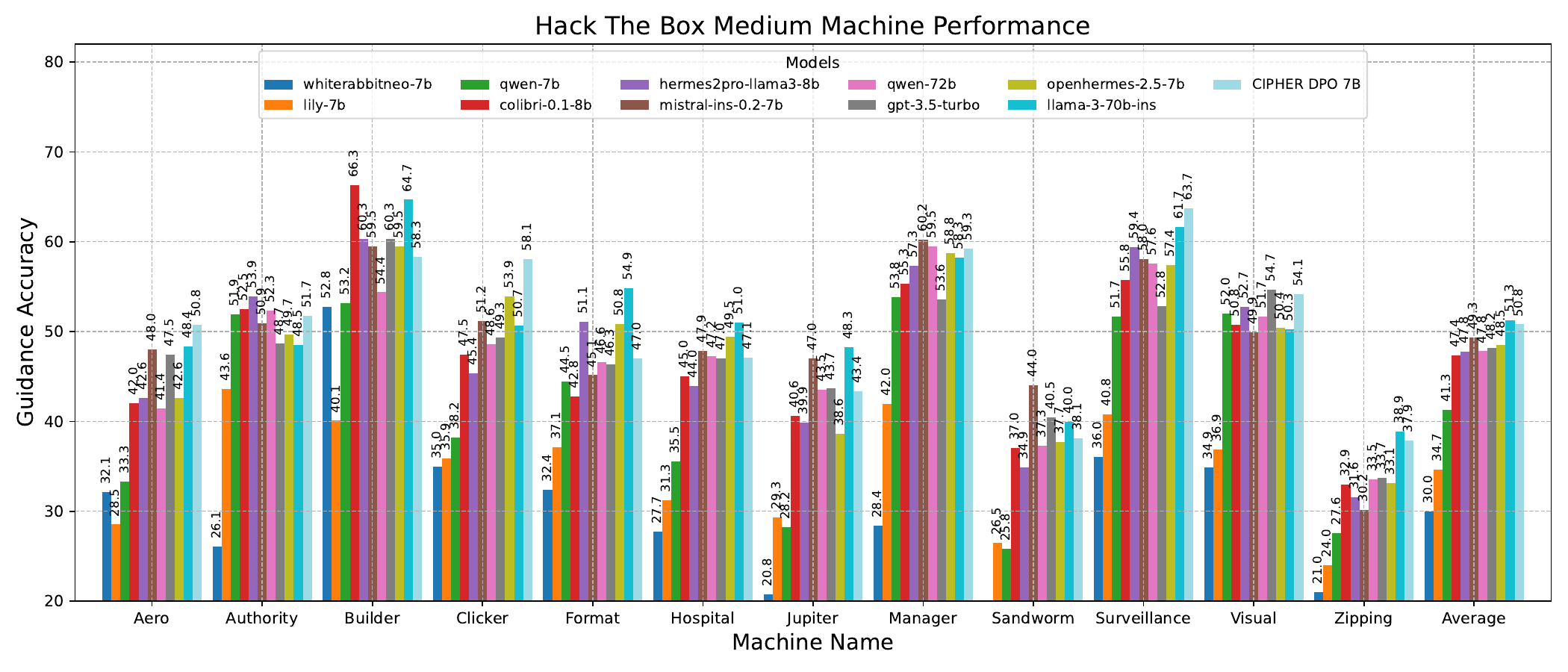}
\end{adjustwidth}
\caption{\textcolor{black}{Medium machine performance results}.}
\label{fig:medium}
\end{figure}

\textcolor{black}{Following this scenario, in the medium machine challenge shown in Figure \ref{fig:medium}, CIPHER experienced a slight performance drop compared to Llama 3 70B. However, despite this small difference, CIPHER still achieved the highest score, creating a notable gap compared to models with 7B parameters. This suggests that CIPHER provides better guidance than the smaller models. It is important to note that Llama 3 70B's high score primarily serves as a benchmark to gauge CIPHER's competitive ability against larger parameter models. Additionally, since Llama 3 70B is used as the evaluation model, its own score may be influenced by bias.}

\begin{figure}[H]
\begin{adjustwidth}{-\extralength}{0cm}
\centering
\includegraphics[width=0.95\linewidth]{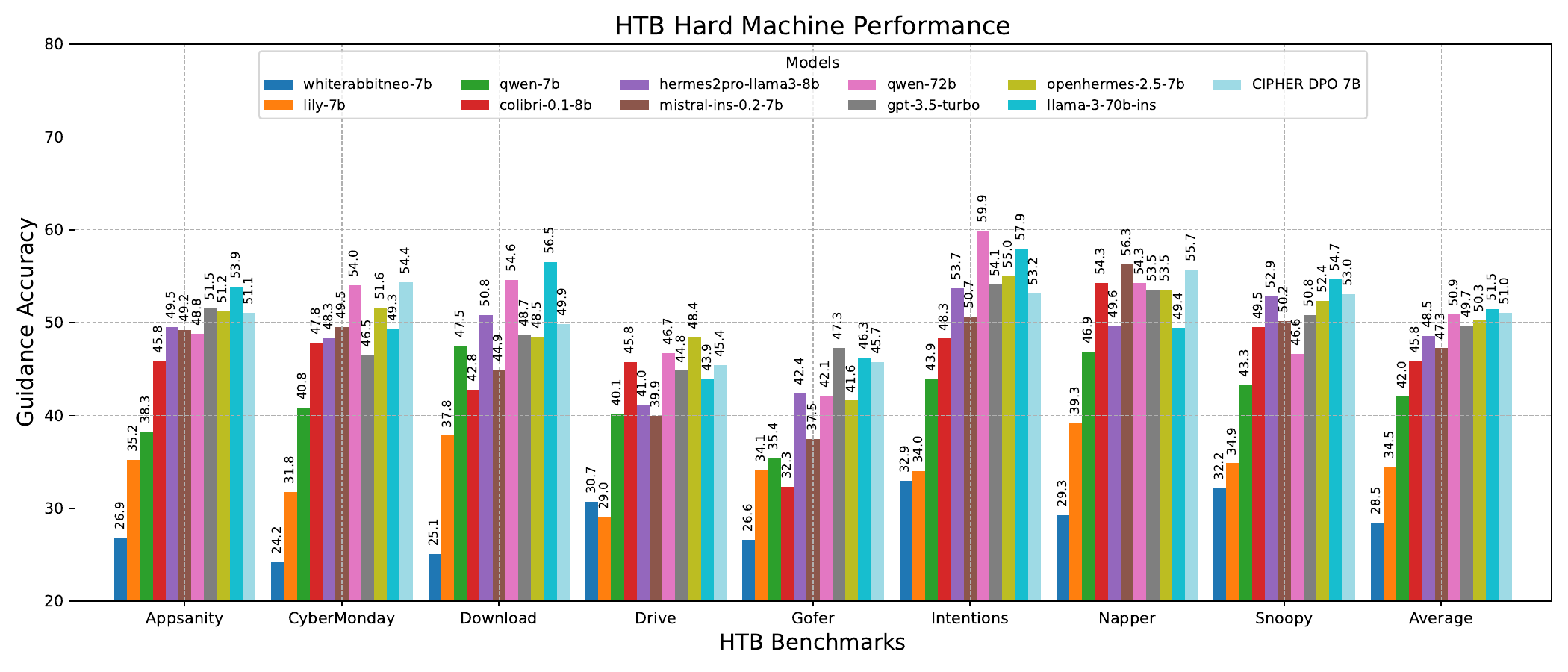}
\end{adjustwidth}
\caption{\textcolor{black}{{Hard machine} 
 performance results}.}
\label{fig:hard}
\end{figure}

\textcolor{black}{This phenomenon is also observed in the hard machine challenge in Figure \ref{fig:hard}, where CIPHER achieves the best score among the 7B models but still falls below the average performance of Llama 3 70B. We believe that, in addition to potential bias, there may be inherent limitations in the 7B model's ability to generalize reasoning and provide accurate guidance. These limitations could potentially be mitigated through dataset scaling and increasing the \mbox{model size}.} 

\begin{figure}[H]
\includegraphics[width=1\linewidth]{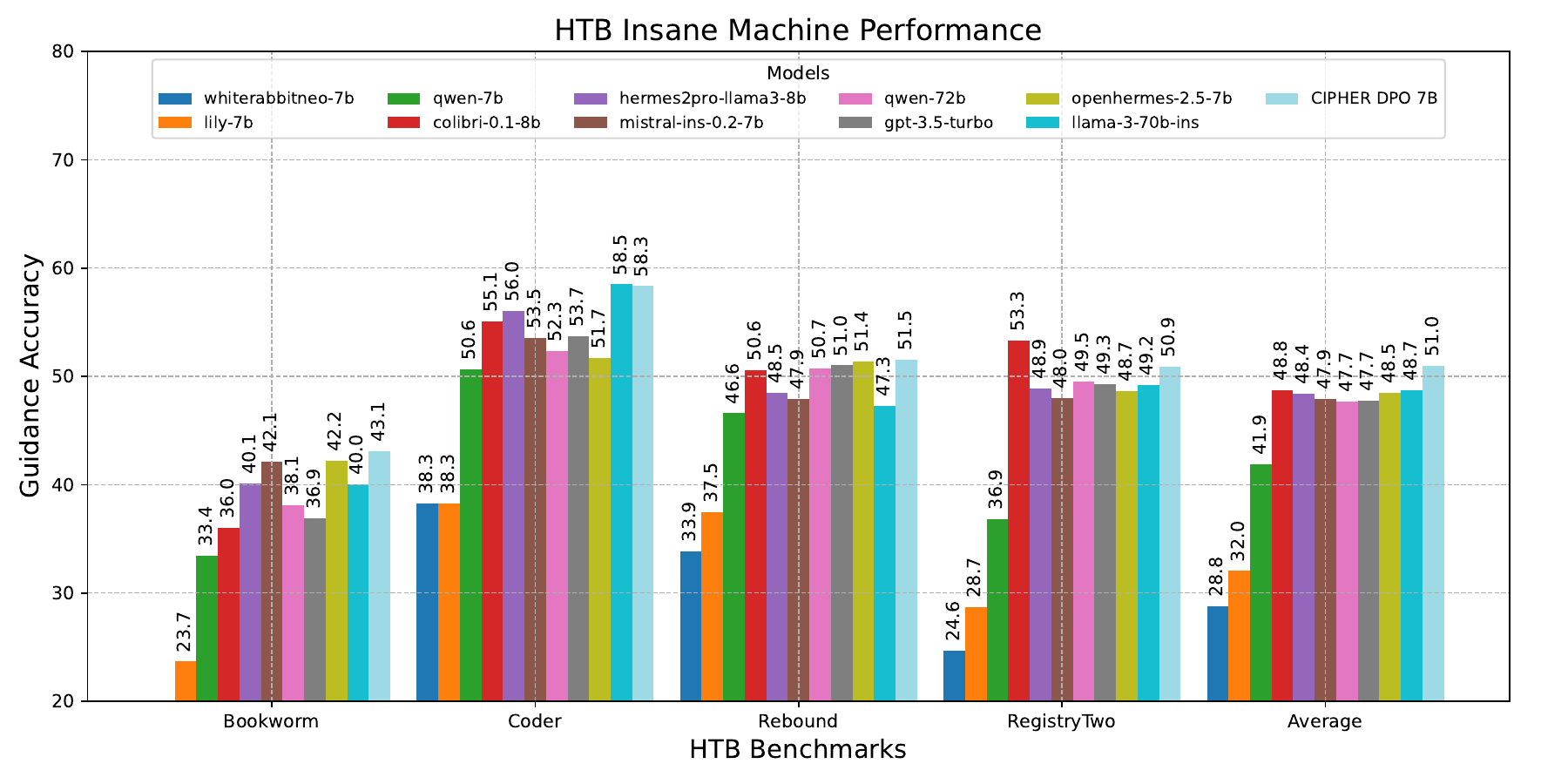}
\caption{\textcolor{black}{{Insane machine} 
 performance results}.}
\label{fig:insane}
\end{figure}

\textcolor{black}{Moving on to the insane difficulty machine's case in Figure \ref{fig:insane}, multiple pieces of information must be combined and a chained exploitation plan is required. Surprisingly, CIPHER leads the benchmark with a 2-point gap over all other models. This demonstrates CIPHER's significant penetration testing guidance capabilities in extremely challenging scenarios that demand out-of-the-box exploitation, where even higher-parameter models reach \mbox{their limits}.}

\vspace{-9pt}

This experiment proves that penetration testing intuition remains a challenge that requires experience rather than just vast knowledge already present in large parameter LLMs. By training CIPHER with a penetration testing-focused guidance dataset, CIPHER can guide users more accurately on unseen vulnerable machines than other specific cybersecurity/penetration testing models and even larger parameter general models.

The results highlight CIPHER's effectiveness across different difficulty levels, particularly excelling in easy and insane categories. This balanced performance across varying complexities underscores the model's versatility and robustness in penetration testing scenarios. The consistent superiority over models with significantly larger parameter counts emphasizes the importance of specialized training and domain-specific knowledge in achieving high-quality results in targeted tasks like penetration testing guidance.

\textcolor{black}{Overall, CIPHER achieved the highest total average score, demonstrating superior performance in both the easy and insane tasks. It also excelled in the medium and hard tasks compared to other pentesting models of its size, placing second only to Llama 3 70B. This may be due to the fact that Llama 3 70B was also the evaluator, and its larger parameter size could have contributed to its advantage. Additionally, we discuss potential improvements for CIPHER in Section \ref{section:scaling}.}

\subsection{PurpleLlama CyberSecEval}

\begin{figure}[H]
\centering
\includegraphics[width=1\linewidth]{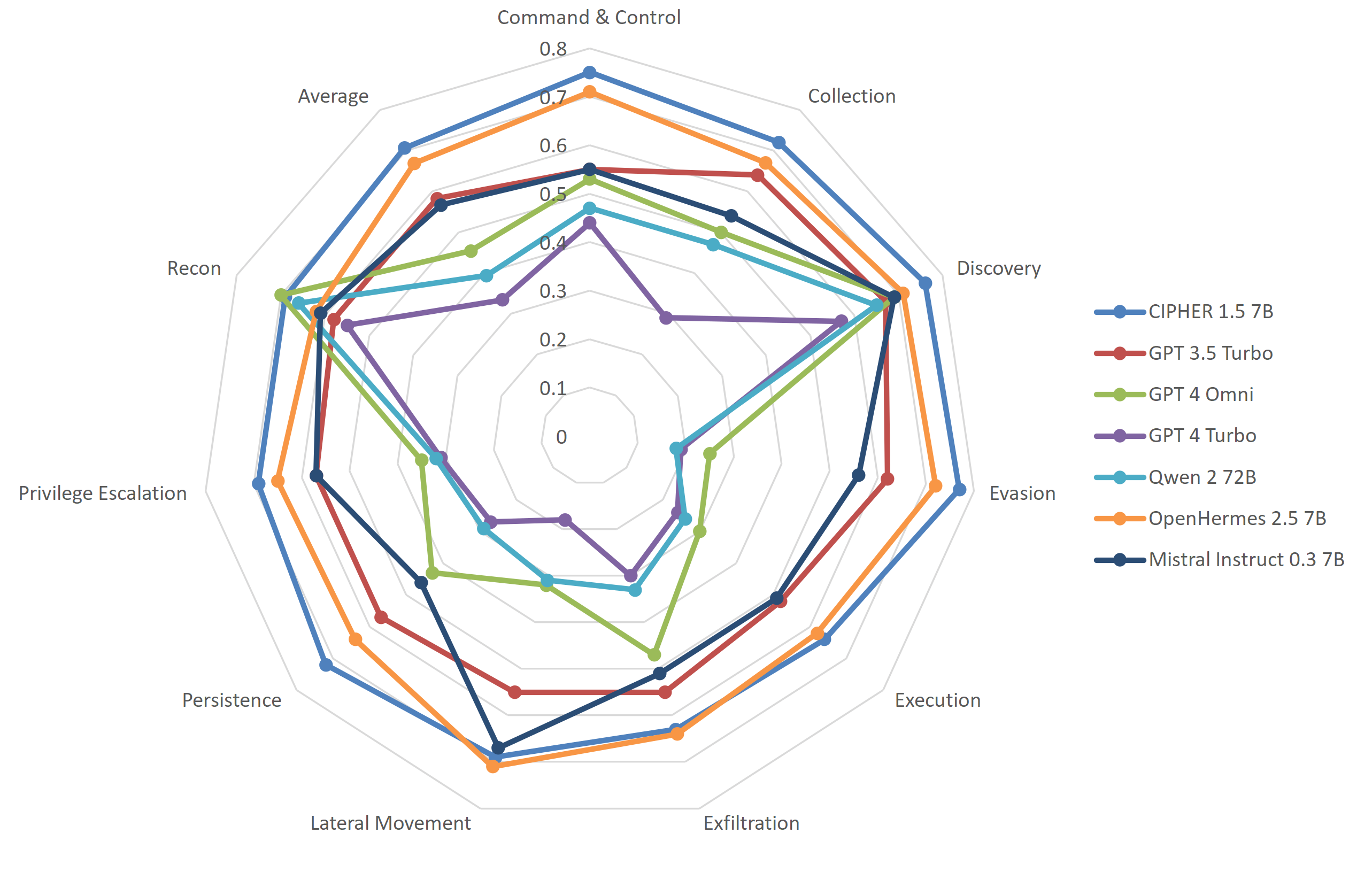} 
\caption{{MITRE ATT\&ACK} 
 capabilities results.}
\label{fig:mitre}
\end{figure}

We also evaluate our model using the most popular cyber security evaluation for LLMs, PurpleLlama CybersecEval2~\cite{bhatt_cyberseceval_2024}. This evaluation was initially designed to measure the safety of the LLM model before being released. Since we are developing a model that accurately suggests offensive exploitation techniques, we can also measure how well our model is capable of attacks listed in MITRE ATT\&CK. The value that we show in {Figure}
 \ref{fig:mitre} is the malicious score, unlike the original paper that showcases the benign score.

CIPHER is based on the Mistral model and leverages the OpenHermes 2.5 dataset for enhanced general knowledge and conversational capabilities. Our synthetic dataset significantly bolsters CIPHER's proficiency in the MITRE ATT\&CK framework, particularly in major categories such as command and control, collection, discovery, evasion, and execution. Given our meticulously curated privilege escalation data, it was anticipated that CIPHER would outperform in persistence and privilege escalation tasks, areas not extensively documented online, which accounts for the subpar performance of other general models. Additionally, the impressive scores of CIPHER are partly because other models incorporate safety measures that limit their ability to engage in offensive activities. In contrast, CIPHER includes these offensive capabilities to ensure its effectiveness in assisting with penetration testing processes, where such skills are critical.

In contrast, we observe that the performance of general models tends to peak in reconnaissance and discovery techniques, even outperforming CIPHER in these areas. This is because reconnaissance and service discovery are common topics that are not considered dangerous. However, using only high-quality augmented write-ups and raw context-hacking techniques material, CIPHER enhances the MITRE ATT\&CK capabilities of its base model and base conversation dataset. While this benchmark does not represent the model's full performance, it showcases its extensive knowledge coverage across different attack techniques.

\subsection{\textcolor{black}{Real-World Beginner Use Case Experiment}}

\textcolor{black}{In this experiment, we explore and analyze how a beginner penetration tester interacts with CIPHER to carry out common tasks such as reconnaissance, vulnerability discovery, and privilege escalation. The goal of this experiment is to evaluate the effectiveness of CIPHER in guiding new penetration testers through the initial stages of a penetration test, while also assessing how well the LLM adapts to novice-level queries and provides appropriate responses. Even though CIPHER has demonstrated strong performance in FARR Flow, this experiment aims to further validate its practical usability in real-world penetration testing scenarios, ensuring that it can handle practical questions effectively.}

\subsubsection{\textcolor{black}{Reconnaissance}}

\begin{figure}[H]
        \includegraphics[width=0.8\linewidth]{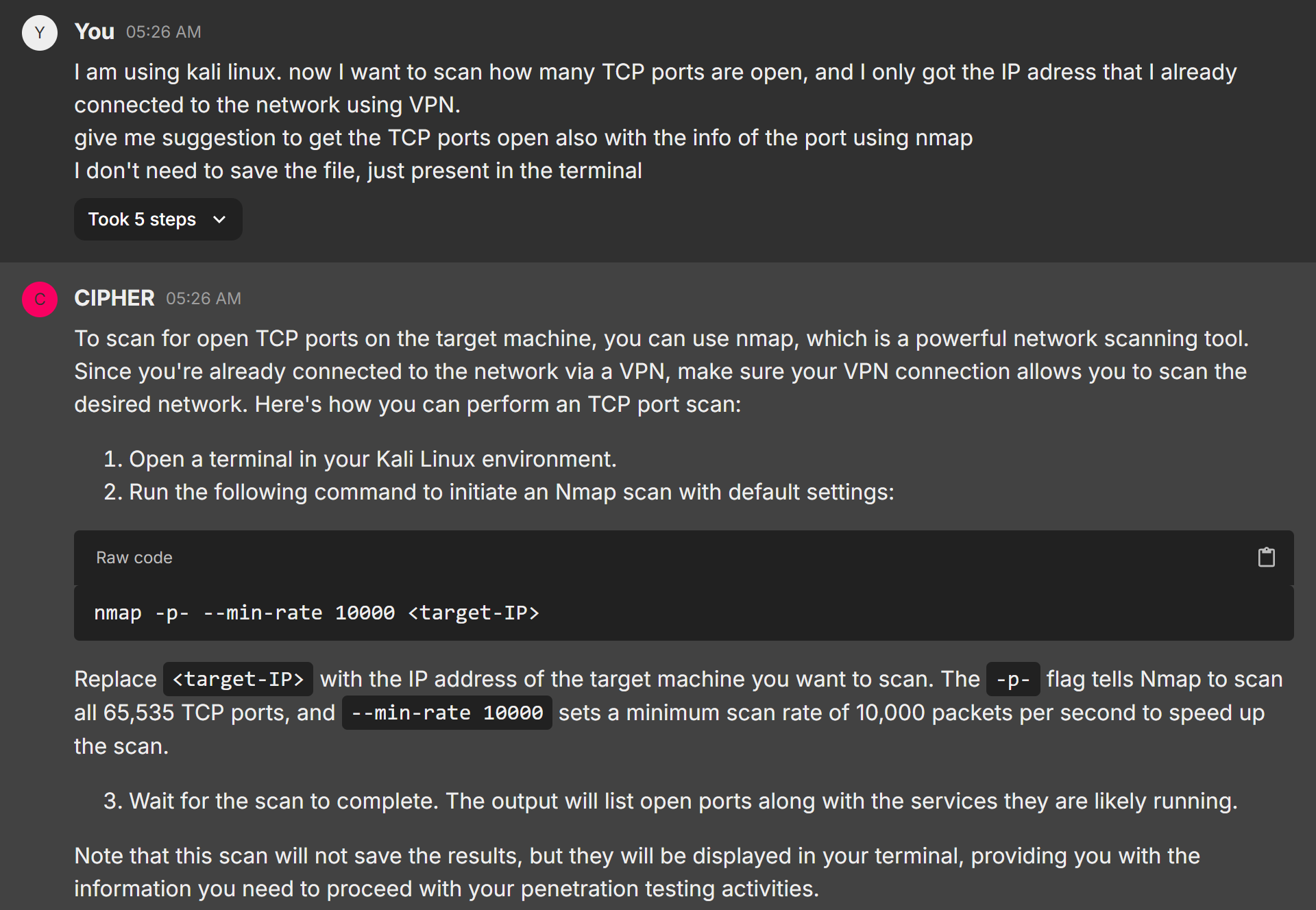}
    \caption{\textcolor{black}{Nmap suggestion to scan all ports.}}
    \label{fig:usage-nmap-recon}
\end{figure}

\textcolor{black}{We begin the experiment with a typical beginner's interaction with CIPHER during a reconnaissance phase. The user is prompted to ask for guidance on the first step in assessing the target system's open ports. Figure \ref{fig:usage-nmap-recon} demonstrates CIPHER's suggestion of using Nmap, a popular network scanning tool, to enumerate open ports on the target system. The model provides the appropriate command syntax \texttt{nmap -p- --min-rate} and also explains its relevance to the overall penetration testing process, emphasizing the need to identify all active services and open ports quickly for further exploitation without saving to any file.}

\begin{figure}[H]
      \includegraphics[width=0.8\linewidth]{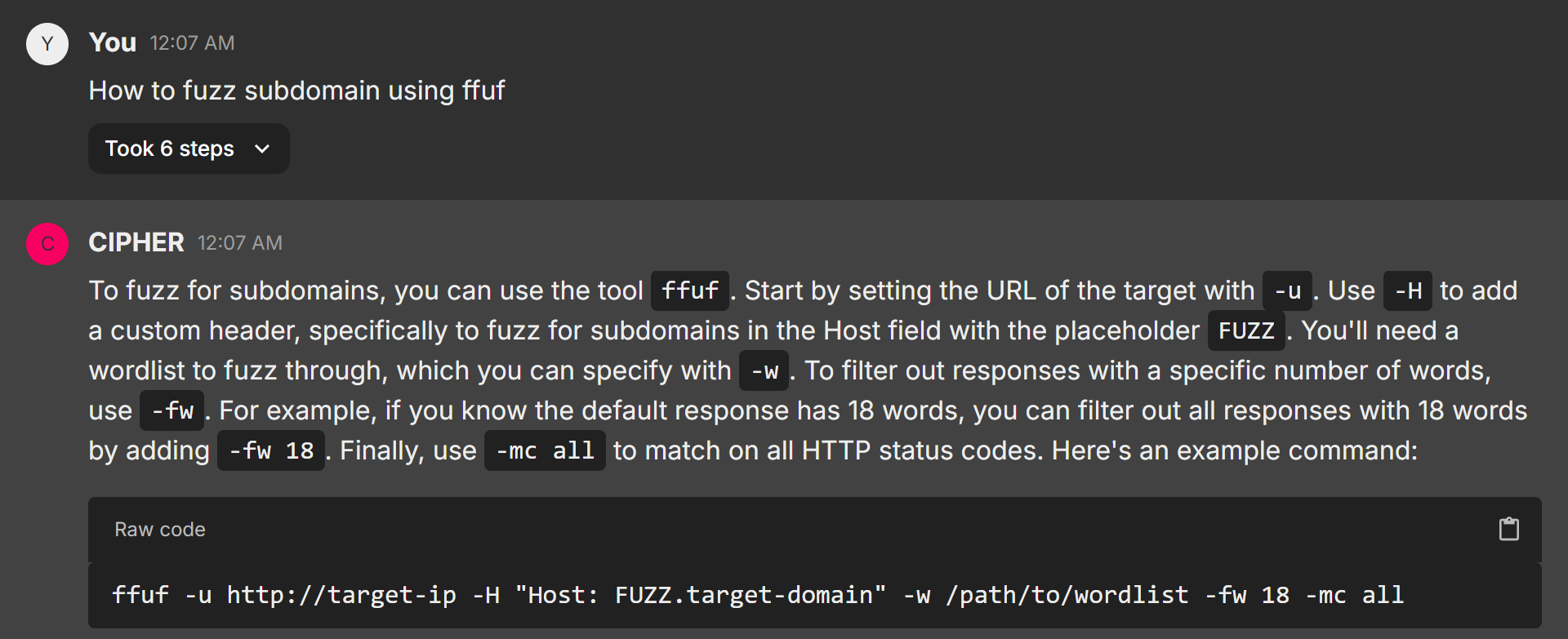}
        \caption{\textcolor{black}{FFUF subdomain fuzzing.}}
        \label{fig:usage-ffuf-subdomain}
        \end{figure}

\textcolor{black}{The next scenario is involved with domain fuzzing, as shown in Figure \ref{fig:usage-ffuf-subdomain}. Providing an actionable guide for using \texttt{ffuf} in subdomain fuzzing, CIPHER effectively breaks down each component of the command. It explains common parameters like \texttt{-u} for the target URL, \texttt{-H} for adding custom headers, and \texttt{-w} for specifying wordlists, showcasing a solid understanding of both the tool and its practical applications. The use of placeholders such as \texttt{FUZZ} in the Host header is a standard fuzzing technique, and the inclusion of \texttt{-fw 18} and \texttt{-mc all} shows a keen awareness of optimizing results and capturing a wide range of \mbox{HTTP responses}.}

\textcolor{black}{The response is highly effective for both novice and intermediate users, balancing simplicity with technical detail. CIPHER's guidance is structured well, leaving room for customization, and demonstrates proficiency in subdomain enumeration. It could be enhanced with tips on handling challenges like rate-limiting and optimizing wordlists, but overall, it successfully conveys core principles and best practices as a penetration \mbox{testing assistant}.}

\begin{figure}[H]
    \includegraphics[width=0.8\linewidth]{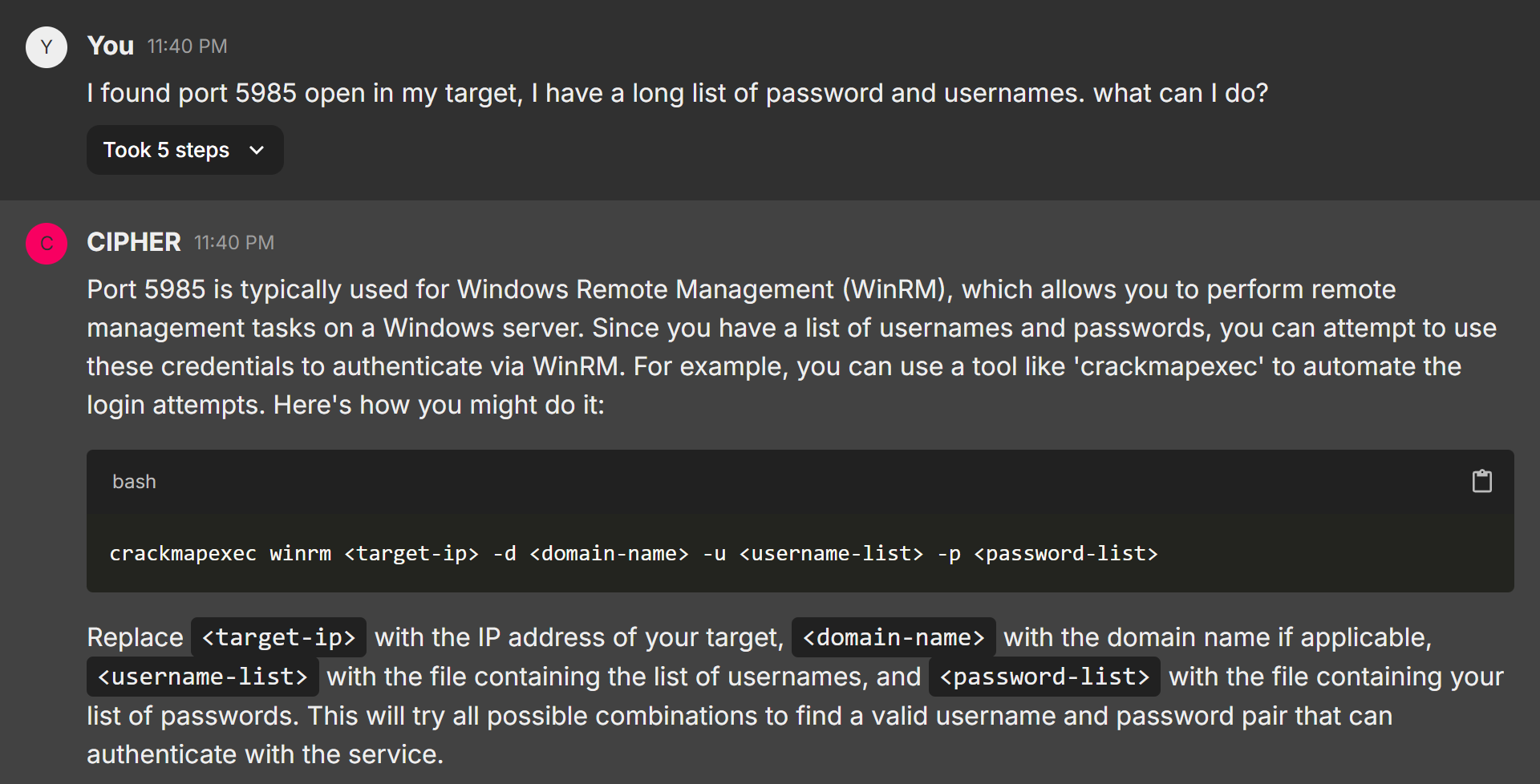}
        \caption{\textcolor{black}{CrackMapExec bruteforcing WinRM service.}}
        \label{fig:usage-cme-brute}
\end{figure}

\textcolor{black}{In the next scenario, the user reports discovering port 5985 open on the target system and asks for advice on utilizing a long list of usernames and passwords. Figure \ref{fig:usage-cme-brute} shows CIPHER’s response, providing a concise explanation of port 5985 being associated with Windows Remote Management (WinRM). CIPHER recommends using \texttt{crackmapexec} to automate testing combinations of usernames and passwords against the WinRM service, guiding the user with clear command syntax: \texttt{crackmapexec winrm <target-ip> -d <domain-name> -u <username-list> -p <password-list>} and explaining each element's role in the process.}

\vspace{-9pt}

\vspace{-9pt}

\textcolor{black}{This guidance is technically sound and user-friendly, offering both a practical tool for automating brute force attempts and explaining the relevance of port 5985. The response includes all the essential details, such as why this method is suitable for WinRM, making it a valuable resource to learn penetration testing. The combination of clear instructions and technical insight highlights CIPHER’s effectiveness as an educational tool in this context.}

\subsubsection{\textcolor{black}{Tools Suggestion}}
\begin{figure}[H]
  \includegraphics[width=0.8\linewidth]{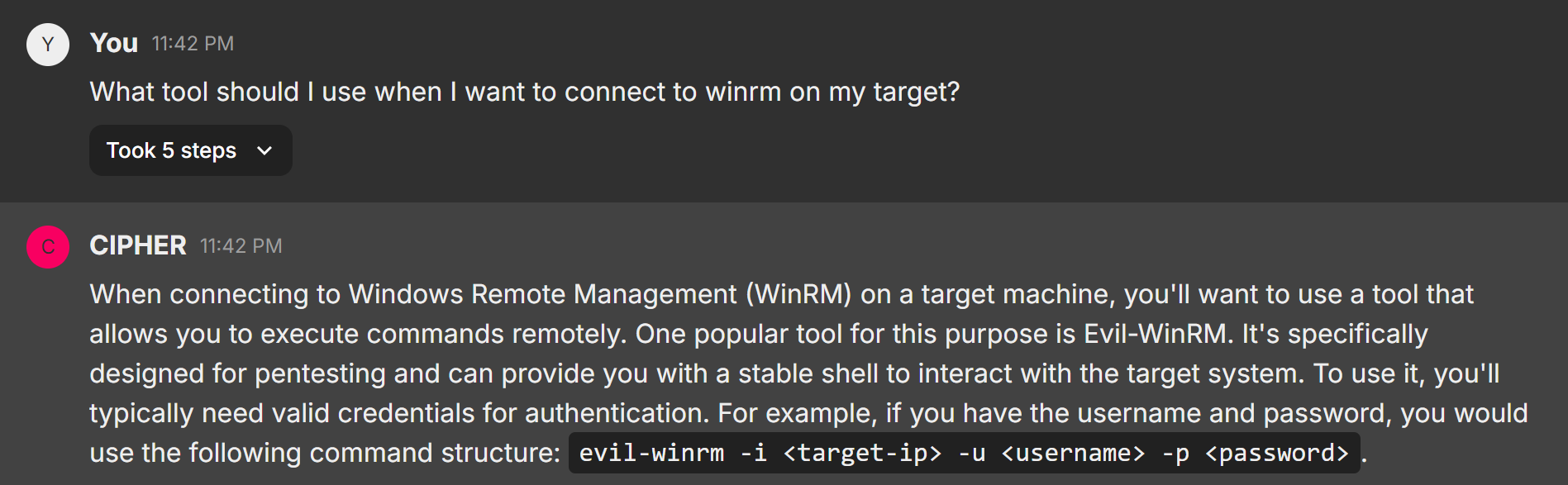}
    \caption{\textcolor{black}{Evil-WinRM as tool suggestion.}}
    \label{fig:usage-ewinrm}
  \end{figure}

\textcolor{black}{In this scenario, the user inquires about the appropriate tool for connecting to WinRM on a target system. Figure \ref{fig:usage-ewinrm} shows CIPHER’s response, recommending the use of Evil-WinRM, a tool specifically designed for penetration testing. CIPHER explains that Evil-WinRM provides a reliable shell for remote command execution and offers clear command syntax: \texttt{evil-winrm -i <target-ip> -u <username> -p <password>}, where the user inputs the target IP, username, and password.}

\textcolor{black}{This interaction demonstrates CIPHER's deep understanding of penetration testing techniques. The model not only suggests the correct tool but also explains its relevance, emphasizing the need for valid credentials when working with WinRM-enabled systems. By providing clear instructions, CIPHER ensures that even novice users can confidently execute the task, making it a valuable educational resource for penetration testers.}

\begin{figure}[H]
      \includegraphics[width=0.8\linewidth]{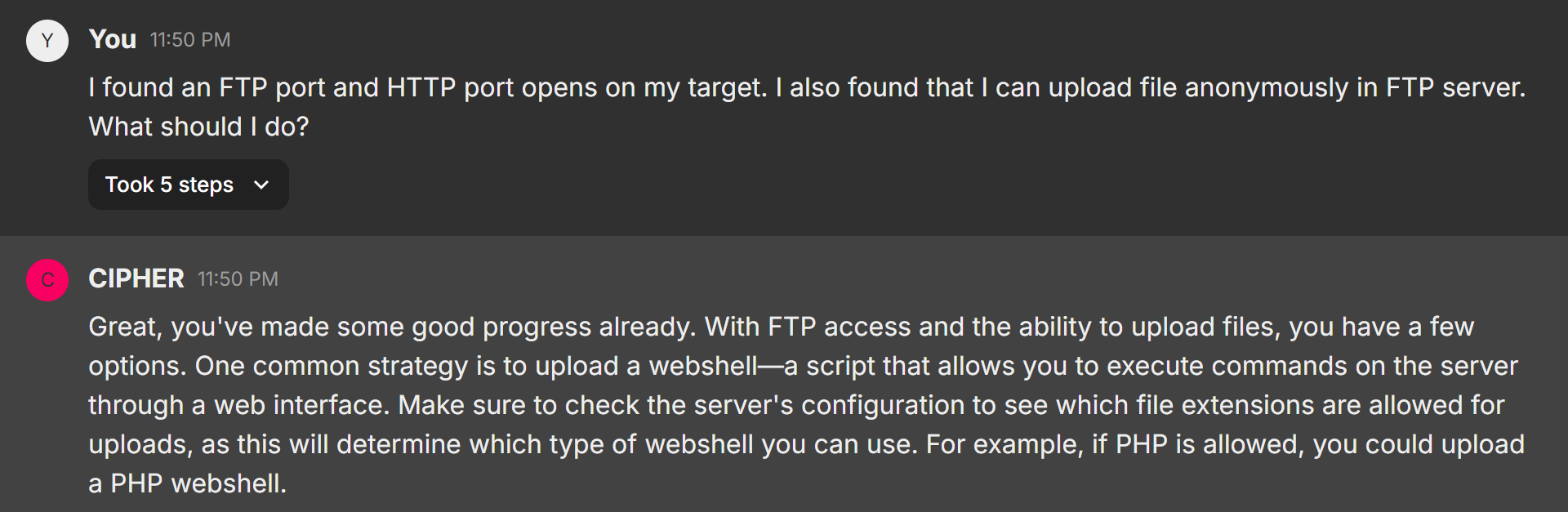}
        \caption{\textcolor{black}{FTP webshell recommendation.}}
        \label{fig:usage-ftp-webshell}
\end{figure}

\textcolor{black}{In this scenario, the user discovers that both FTP and HTTP ports are open on the target, with the ability to upload files anonymously to the FTP server. Figure \ref{fig:usage-ftp-webshell} shows CIPHER’s response, which acknowledges the progress and suggests a common strategy: uploading a webshell. This is a script that allows command execution on the server via a web interface. CIPHER advises the user to first check the server's configuration to determine which file extensions are allowed for uploads, as this will dictate the type of webshell that can be used. For example, if the server supports PHP, the user could upload a PHP webshell to gain control over the system.}

\textcolor{black}{CIPHER's guidance shows an understanding of the penetration testing process, particularly when dealing with misconfigured FTP and HTTP services. The suggestion to check file extension permissions reflects the nuanced approach needed when attempting to escalate access, ensuring that the webshell method is tailored to the target’s configuration.}

\begin{figure}[H]
        \includegraphics[width=0.8\linewidth]{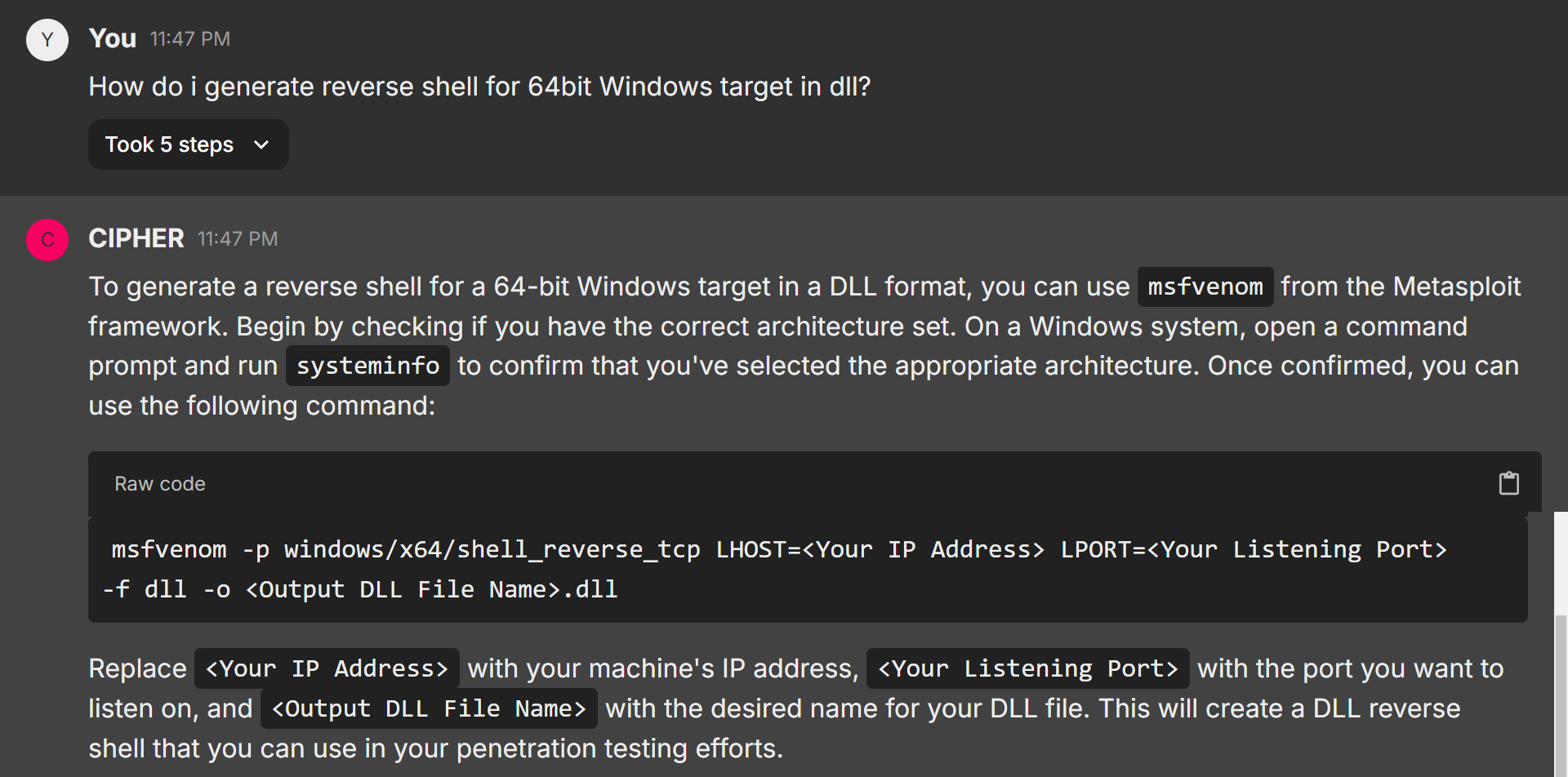}
    \caption{\textcolor{black}{Exploit creation suggestion}.}
    \label{fig:usage-msfvenom}
\end{figure}

\textcolor{black}{In the next case study, the user asks how to generate a reverse shell for a 64-bit Windows target in a DLL format. Figure \ref{fig:usage-msfvenom} illustrates CIPHER's detailed response, recommending the use of msfvenom from the Metasploit framework. CIPHER begins by advising the user to check the system's architecture using the systeminfo command, ensuring the correct architecture set is used. Once confirmed, CIPHER provides a command syntax to generate the DLL reverse shell. The user is instructed to replace placeholders with appropriate values to create the reverse shell, which can then be used in the penetration testing process.}

\textcolor{black}{This response reflects a solid understanding of reverse shell generation, covering both technical accuracy and practical guidance. CIPHER explains not only how to generate the reverse shell but also emphasizes verifying the target’s architecture, demonstrating a nuanced approach to penetration testing and ensuring that the process runs smoothly.}

\subsubsection{\textcolor{black}{Attack Vector Suggestion}}

\begin{figure}[H]
        \includegraphics[width=0.8\linewidth]{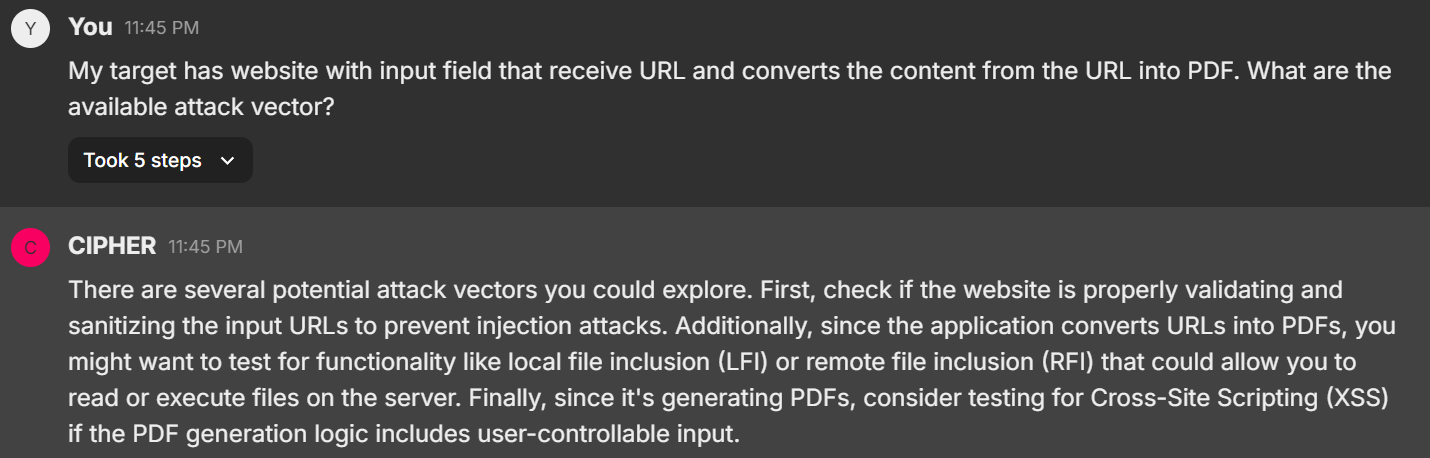}
    \caption{\textcolor{black}{Attack vector suggestion}.}
    \label{fig:usage-attack-verctor}
\end{figure}

\textcolor{black}{In this scenario, the user asks about potential attack vectors for a website that accepts URLs and converts the content into PDF format. Figure \ref{fig:usage-attack-verctor} displays CIPHER’s response, which suggests several potential attack vectors, starting with Server-Side Request Forgery (SSRF). This attack allows the attacker to manipulate the server into fetching resources from unintended locations, possibly exposing internal services or sensitive data. CIPHER also points out the security risks associated with the PDF generation process, which could be exploited via Cross-Site Scripting (XSS) or SQL Injection if not properly secured. The model emphasizes the importance of gathering as much information as possible about the target system through enumeration before launching any attacks, demonstrating a nuanced understanding of web-based vulnerabilities.}

\begin{figure}[H]
        \includegraphics[width=0.73\linewidth]{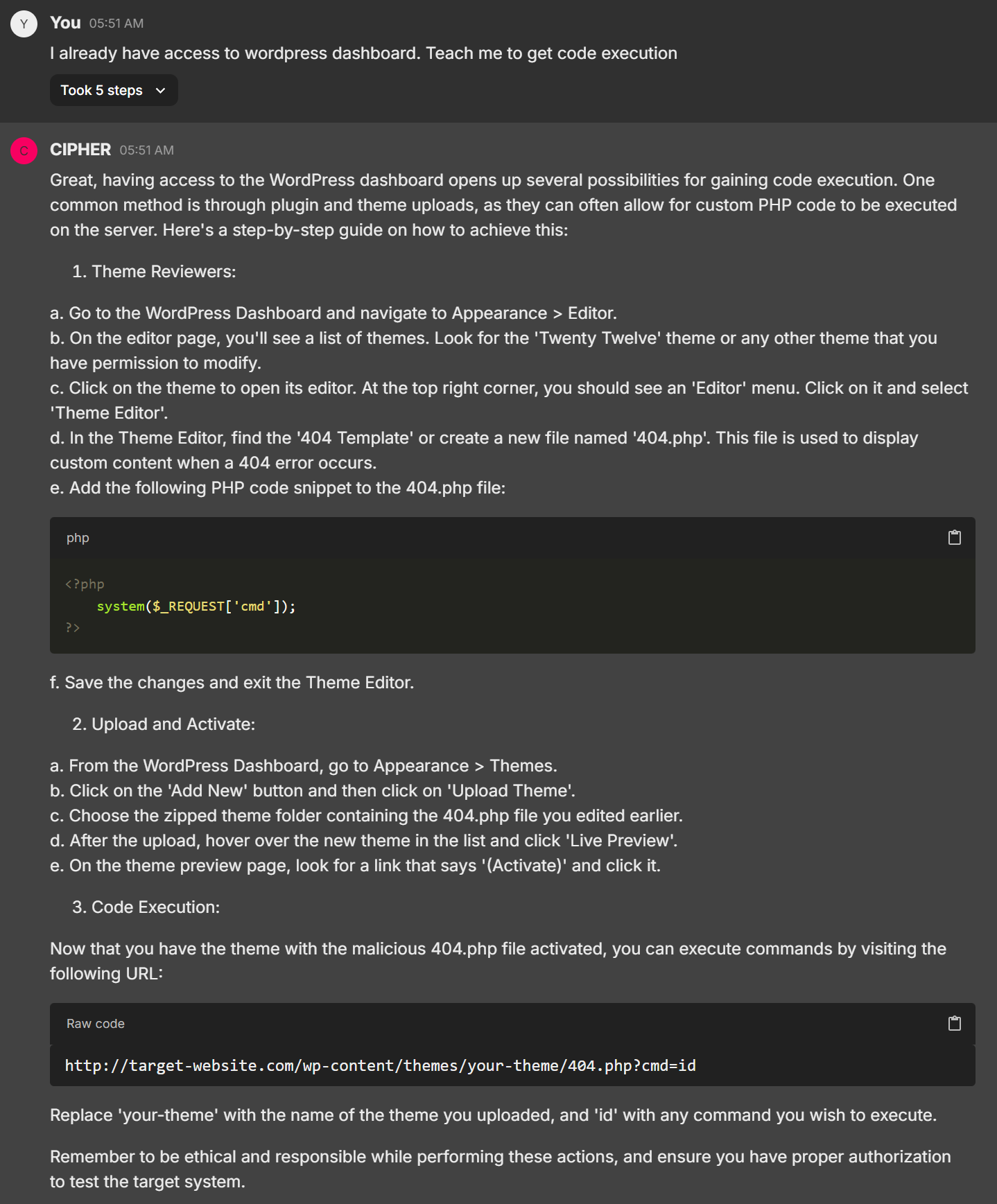}
    \caption{\textcolor{black}{Wordpress RCE instruction}.}
    \label{fig:usage-wp-rce}
    \end{figure}

\textcolor{black}{CIPHER's response in Figure \ref{fig:usage-wp-rce} provides a clear and practical guide to gaining code execution via WordPress, offering step-by-step instructions on file manipulation, such as editing the `404.php' file and uploading a theme. The attention to detail ensures that users of all experience levels can follow along easily, with the model demonstrating a solid understanding of WordPress-specific vulnerabilities and how themes can be exploited for remote code execution.}

    \textcolor{black}{The choice to use a \texttt{404.php} file and inject the \texttt{system(\$\_REQUEST['cmd']);} payload shows smart tactics, as this file is often overlooked. While the response could enhance stealth advice, such as clearing logs or minimizing site disruption, it effectively balances clarity and technical depth, making it suitable for both beginners and more advanced users. CIPHER's output showcases a well-rounded understanding of penetration testing within WordPress environments.}

\subsubsection{\textcolor{black}{Privilege Escalation Suggestion}}

\begin{figure}[H]
        \includegraphics[width=0.8\linewidth]{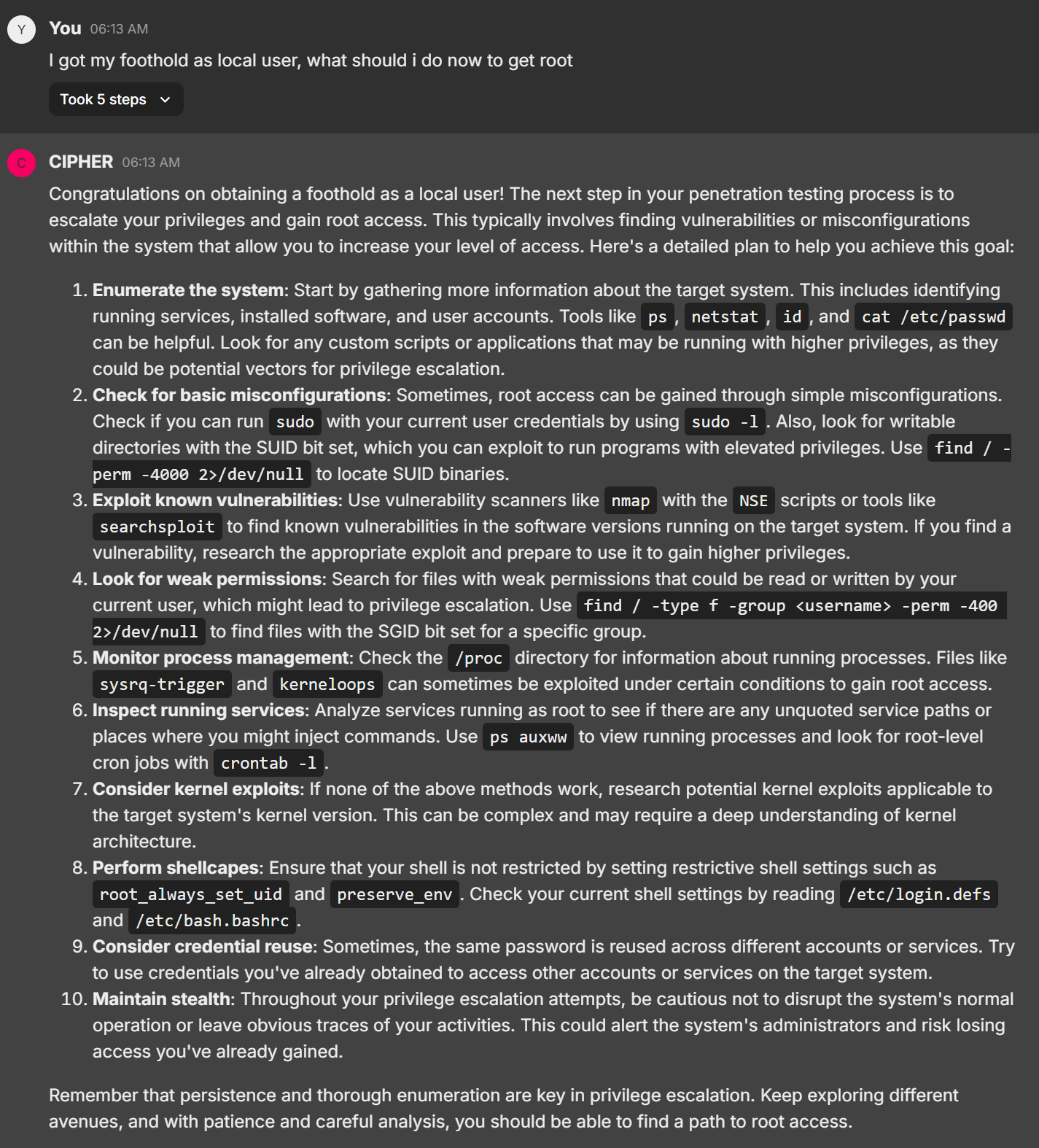}
    \caption{\textcolor{black}{Privilege escalation advice after foothold first gained}.}
    \label{fig:usage-pe-advice}
\end{figure}

\textcolor{black}{Shifting to the next example response in Figure \ref{fig:usage-pe-advice}, CIPHER offers a comprehensive, systematic approach to privilege escalation after gaining a foothold as a local user, aligning with established penetration testing methodologies. Starting with enumeration using tools like `ps', `netstat', `id', and `sudo -l', CIPHER ensures the user collects key system information, laying a strong foundation for identifying privilege escalation opportunities. The explanation of SUID/SGID files and how to locate them with `find' is particularly valuable, providing actionable advice for both novice and experienced users.}

\textcolor{black}{CIPHER further demonstrates its depth by covering misconfigurations, kernel exploits, and shell escape techniques, which are crucial for exploiting system weaknesses. The model's attention to advanced scenarios, such as credential reuse and inspecting insecure cron jobs, along with its focus on maintaining stealth throughout the process, shows a strong understanding of real-world penetration testing tactics. This balance of clarity and technical depth makes CIPHER’s guidance highly effective for beginners.}

\begin{figure}[H]
       \includegraphics[width=0.8\linewidth]{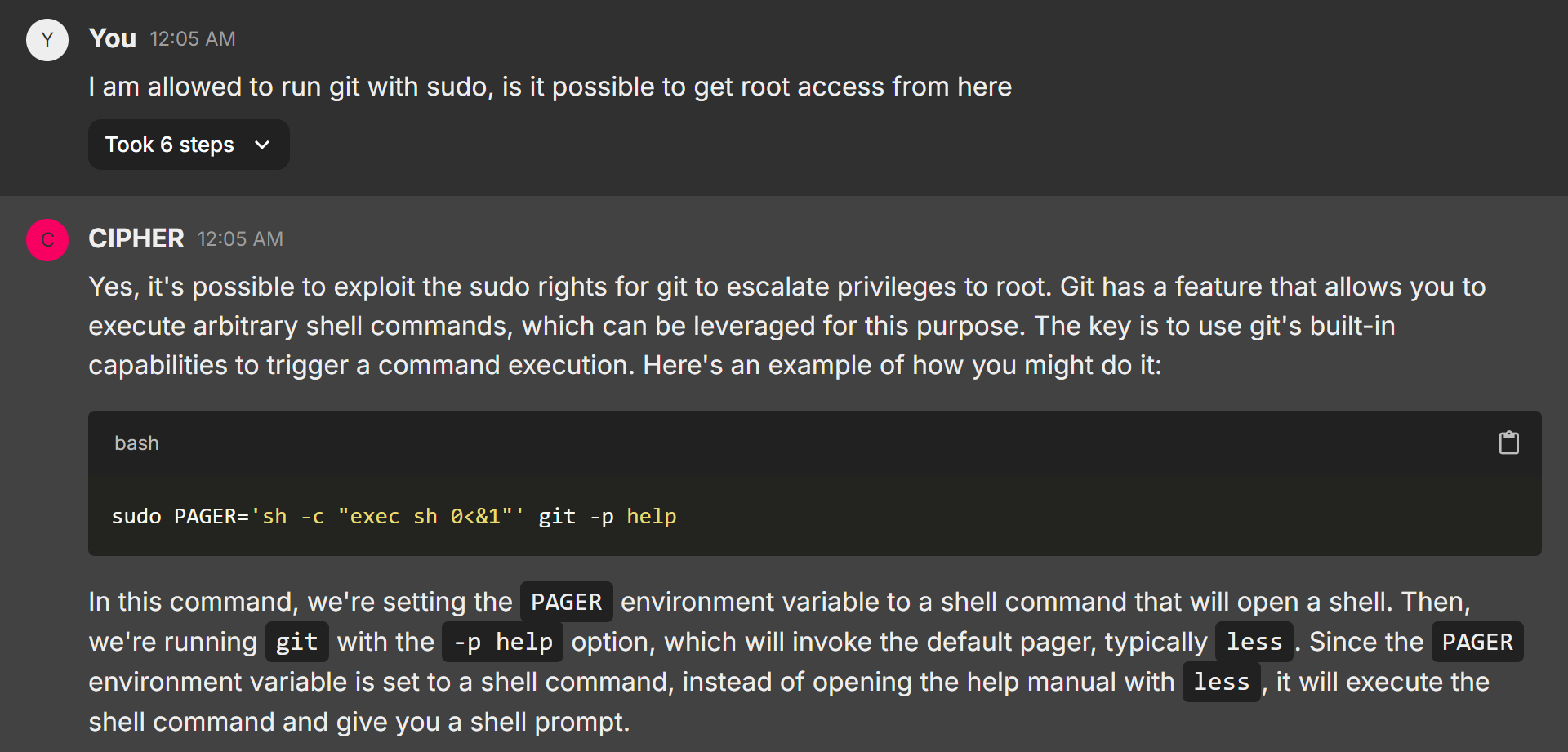}
    \caption{\textcolor{black}{Privilige escalation using misconfigured Git}.}
    \label{fig:usage-git-pe}
\end{figure}

\textcolor{black}{The next study in Figure \ref{fig:usage-git-pe} showcases CIPHER's solid understanding of privilege escalation by helping a user exploiting Git's `PAGER` environment variable to execute arbitrary shell commands. The command \texttt{sudo PAGER='sh -c ``exec sh 0<\&1''' git -p help} takes advantage of Git's behavior to spawn a shell, effectively bypassing restricted environments. CIPHER explains each component of the command clearly, ensuring that even users unfamiliar with environment variables can follow along, while demonstrating technical depth and practicality.}

\vspace{-9pt}

\textcolor{black}{The response balances usability and insight, making it highly accessible to beginners. CIPHER’s guidance is clear, actionable, and well-aligned with established privilege escalation techniques. A minor enhancement could include addressing potential limitations in more restrictive environments, but overall, the quality of CIPHER’s output is excellent, effectively supporting penetration testers to gain highest privilege.}

\textcolor{black}{Based on this experiment, we have proven that CIPHER can be effectively used as a practical penetration testing assistant, guiding users through the utilization of multiple tools from reconnaissance to post-exploitation. We observed that while CIPHER performs well overall, there are areas for improvement, such as providing more detailed explanations, example cases, or alternative commands. These improvements can be achieved by scaling up the dataset to support longer and more detailed responses. There is also available room for improvement in model size; despite this, CIPHER is able to respond accurately and assist users effectively.}

\section{Discussion and Future Works}
\subsection{Limitations}
While CIPHER demonstrates superior average performance compared to other models, including the higher parameter Llama 3 70B, in the FARR Flow reasoning evaluation, this does not imply that CIPHER excels at reasoning during problem debugging. Our observations indicate that CIPHER is proficient at predicting the accurate next step given limited information. However, acting often requires more detailed information, which beginners might need to ask for. CIPHER is not capable of troubleshooting the commands it provides. It sometimes makes mistakes, such as hallucinating command arguments. It tends to hallucinate even more when further questions about the correct command \mbox{are asked}.

We believe this issue arises from a bias in the penetration testing data. When a user reports a problem, CIPHER relates it to penetration testing. For instance, if a provided penetration testing command does not work, instead of correcting the command or asking for the arguments used, CIPHER might assume the target is unreachable or suggest alternative tools for the same purpose. This focus degrades its ability to handle general command errors or Linux command troubleshooting despite incorporating raw datasets that include command line documentation.

Since CIPHER is designed to understand the current situation and suggest subsequent actions, we did not heavily emphasize coding datasets. CIPHER's coding performance can be enhanced by changing the base model to a coding model or using a higher-parameter base model. From a data-centric perspective, improving coding troubleshooting performance by creating synthetic datasets is possible. This can be achieved similarly to the CIPHER method, involving experts to fix code and provide explanations, reasoning, \mbox{and~examples.}

Our FARR Flow augmentation is imperfect; some generated augmentations still mention ``the author'' when they are supposed to be just like finding notes. We have explicitly forbidden this behavior, but it is still in some samples. This could be caused by the limitation of the model used for generation. However, this does not affect the model response when augmented into a question, and even if it exists, it will only cause minor issues in perspective sentences.

Another difficulty in FARR Flow augmentation is providing challenging findings. Generated findings often contain hints, such as ``found a zip file, need to crack the password''. While this challenges the model to determine which tools to use for cracking, it already suggests a line of reasoning. As a result, the model's focus will be on cracking the password instead of first checking the metadata or trying known target user passwords. This nuance needs to be understood by the model generating the FARR Flow. Although this is not a direct limitation of FARR Flow augmentation, it opens up future work to refine the pipeline and eliminate such results. Rather than waiting for AGI-level LLM, a feasible approach is to use agents or more advanced chain algorithms.

In the FARR Flow evaluation, we use Llama 3 70B as the judge LLM. Even though it is a state-of-the-art model, it has limitations and biases. Different judges can introduce different biases. Using multiple judge LLMs in future work is a possible solution to reduce this bias.

\subsection{Scaling}\label{section:scaling}
CIPHER's current performance in penetration testing is influenced by the size of its dataset and model. There are several strategies to enhance the model's capabilities further.

The first strategy involves scaling the dataset. 
This can be improved by incorporating more diverse and specific augmentations for the same dataset, using techniques like Orca~\cite{mukherjee_orca_2023} and Evolve instruct~\cite{xu2023wizardlmempoweringlargelanguage}, allowing the model to generalize better. \textcolor{black}{Also, more contextual details can be added into the dataset for the model to understand details of the current chunk to be generated such as date, scope of the context, or even source metadata, preventing the model hallucinating on unknown context. In addition, wide coverage of penetration testing knowledge can be obtained by adding CVE, CWE, and CAPEC as resources for CIPHER dataset augmentation.} Currently, the cost of CIPHER dataset augmentation ranges from USD 800 
to USD 1000, excluding experimental costs, which can reach around USD 10,000 for full augmentation with the current dataset coverage.

The second strategy is to increase the model size. The current 7B model is insufficient for handling complex tasks like real penetration testing. Larger models with more parameters can significantly enhance generalization ability~\cite{zhang2021understanding}. Training CIPHER to its best performance at 7 epochs took approximately 4 days on 8xH100 GPUs with a 1M dataset. Training a larger model with 70B parameters, such as Llama 3 or Qwen1.5, is estimated to take about one month or more.
There is considerable potential to improve performance, and with sufficient research, we are confident that LLMs can achieve expert capabilities in assisting with penetration testing tasks.

\subsection{Measuring Other Penetration Testing Capabilities} \label{improvebenchmark}
Our FARR Flow augmentation method extracts a raw list of critical information from penetration testing write-ups. While our research primarily focuses on measuring the accuracy of next-step prediction guidance, it is possible to craft different question structures from the four items provided in FARR Flow steps. This approach can be used to assess various capabilities, such as guessing the findings when given an action or predicting the outcome when shown the reasoning.

An exciting way to evaluate the model's reasoning capabilities is by not providing the next findings if the model's action is incorrect. In this scenario, at the initial step, the model is given only the target's IP address. We then continuously prompt the model to generate the following action. If the action likely results in the same outcome—verified by targeting the same service and vulnerability, even with different tools—the evaluator provides the results as the next piece of information.

Using this method, we can measure how quickly the model helps the user penetrate and take over the target machine. A model that requires fewer actions to produce results and can gain privileged access more efficiently will be deemed superior.

\textcolor{black}{For future work, we plan to deeply explore how CIPHER can assist in uncontrolled environments, where system behaviors are unpredictable or information is incomplete. This exploration will allow us to assess CIPHER's ability to adapt to irregular scenarios, such as misconfigured services or unexpected vulnerabilities. By comparing its performance in both controlled and uncontrolled environments, we aim to evaluate how well CIPHER can identify alternative actions and adapt its predictions when faced with less predictable conditions. This investigation will provide valuable insights into the robustness and potential of CIPHER's integration into FARR Flow's next-step prediction accuracy testing.}

\subsection{Ethical Consideration}

We have decided not to release the CIPHER model due to ethical considerations. However, to encourage research and improvement in the domain of penetration testing with LLMs, we will publicly release our novel FARR Flow augmentation and evaluation frameworks on GitHub after our paper is published. We hope that the community will develop more sophisticated and standardized automatic pentesting LLM evaluation methods.

\section{Conclusions}

\textcolor{black}{Our research focused on developing CIPHER, an AI chatbot designed to guide beginner penetration testers through the pentesting process. We utilized synthetic novice and expert conversations, high-quality documentation, and open-source task datasets. Synthetic questions were crafted to assess the model's understanding and ability to suggest accurate actions.}

\textcolor{black}{Our novel FARR Flow evaluation revealed that CIPHER slightly decreases general capabilities but significantly enhances pentesting skills, providing more accurate suggestions than other models on average, particularly in the easy and insane machine categories. However, it shows slight underperformance in the medium and hard categories, but still leads among the 7B parameter models.}

\textcolor{black}{In practical testing, CIPHER demonstrated it can assist effectively with its current parameter size, handling a wide range of tasks from reconnaissance to post-exploitation. There is room for improvement, particularly in providing more detailed explanations and alternative actions.}

\textcolor{black}{Future enhancements could involve full dataset augmentation, additional security knowledge base, and larger parameter models, as well as refining the FARR Flow method by crafting more complex questions or developing a virtual vulnerable machine LLM evaluation based on FARR Flow steps.}

\vspace{6pt} 




\authorcontributions{Conceptualization, D.P.; methodology, D.P.; software, D.P., N.S. and A.A.A.; validation, D.P., N.S., A.A.A., T.-T.-H.L., M.I. and A.Y.K.; formal analysis, D.P., N.S., A.A.A., \mbox{T.-T.-H.L.} and A.Y.K.; investigation, D.P.; resources, D.P.; data curation, D.P.; writing---original draft preparation, D.P.; writing---review and editing, D.P., T.-T.-H.L., A.A.A., N.S. and A.Y.K.; visualization, D.P., \mbox{T.-T.-H.L.} and A.A.A.; supervision, T.-T.-H.L. and H.K.; project administration, H.K.; funding acquisition, H.K. All authors have read and agreed to the published version of the manuscript.}

\funding{{This research was supported} 
 by the MSIT (Ministry of Science and ICT), Korea, under the Convergence Security Core Talent Training Business (Pusan National University) support program (No. RS-2022-II221201) and  was supported by the ITRC (Information Technology Research Center) support program (IITP-2024-RS-2020-II201797), both supervised by the IITP (Institute for Information \& Communications Technology Planning \& Evaluation).}

\institutionalreview{{Not applicable.} 
}

\informedconsent{Not applicable.}

\dataavailability{The data and benchmark presented in this study are available publicly at the our repository (\url{https://github.com/ibndias/CIPHER} accessed on 21 October 2024).}

\conflictsofinterest{The authors declare no conflicts of interest.} 



\begin{adjustwidth}{-\extralength}{0cm}

\reftitle{{References}}
\PublishersNote{}
\end{adjustwidth}
\end{document}